\documentclass[fleqn,usenatbib]{mnras}

\usepackage{newtxtext,newtxmath}

\usepackage[T1]{fontenc}
\usepackage{ae,aecompl}

\usepackage{graphicx}	
\usepackage{amsmath}	
\usepackage{amssymb}	
\usepackage{xspace}
\usepackage{gensymb}

\defcitealias{Poggianti2017a}{Paper I}
\defcitealias{Bellhouse2017}{Paper II}
\defcitealias{Fritz2017}{Paper III}
\defcitealias{Gullieuszik2017}{Paper IV}
\defcitealias{Moretti2018}{Paper V}
\defcitealias{Vulcani2018}{Paper VII}
\defcitealias{Vulcani2017c}{Paper VIII}
\defcitealias{Jaffe2018}{Paper IX}
\defcitealias{Vulcani2018b}{Paper XII}
\defcitealias{Poggianti2019}{Paper XIII}
\defcitealias{Vulcani2018c}{Paper XIV}

\newcommand{\Ha}{H$\alpha$\xspace}

\newcommand{\kube}{{\sc kubeviz}\xspace}
\newcommand{\sinopsis}{{\sc sinopsis}\xspace}
\newcommand{\ma}{$\rm M_\ast$\xspace}
\newcommand{\ms}{$\rm M_\odot$\xspace}
\newcommand{\Ssfr}{$\rm \Sigma_{SFR}$\xspace}
\newcommand{\Sm}{$\rm \Sigma_\ast$\xspace}
\newcommand{\Stg}{$\rm \Sigma_{tot \, gas}$\xspace}

\newcommand{\msk}{$\rm M_\odot \, kpc^{-2}$\xspace}
\newcommand{\msyk}{$\rm M_\odot \, yr^{-1} \, kpc^{-2}$\xspace}

\title[Spatially resolved SFR-Mass relation in local spiral galaxies]{GASP. XX. 
From the { loose} spatially-resolved to the { tight} global SFR-Mass relation in local spiral galaxies}

\author[B. Vulcani et al.]
{Benedetta Vulcani,$^{1}$\thanks{E-mail: benedetta.vulcani@inaf.it (BV)}
Bianca M. Poggianti,$^{1}$
Alessia Moretti,$^{1}$
Andrea Franchetto,$^{2,1}$
\newauthor
Marco Gullieuszik,$^{1}$
Jacopo Fritz,$^{3}$
Daniela Bettoni,$^{1}$
Stephanie Tonnesen,$^{4}$
\newauthor
Mario Radovich,$^{1}$
Yara L. Jaff\'e,$^{5}$
Sean McGee,$^{6}$
Callum Bellhouse,$^{6}$
Giovanni Fasano$^{1}$\\
$^{1}$INAF- Osservatorio astronomico di Padova, Vicolo Osservatorio 5, IT-35122 Padova, Italy\\
$^{2}$Dipartimento di Fisica \& Astronomia ``Galileo Galilei'', Universit\`a di Padova, vicolo dell' Osservatorio 3, IT 35122, Padova, Italy\\
$^{3}$Instituto de Radioastronom\'ia y Astrof\'isica, UNAM, Campus Morelia, A.P. 3-72, C.P. 58089, Mexico\\
$^{4}$Center for Computational Astrophysics, Flatiron Institute, 162 5th Ave, New York, NY 10010, USA\\
$^{5}$Instituto de F\'isica y Astronom\'ia, Universidad de Valpara\'iso, Avda. Gran Breta\~na 1111 Valpara\'iso, Chile\\
$^{6}$University of Birmingham School of Physics and Astronomy, Edgbaston, Birmingham, United Kingdom}

\date{Accepted XXX. Received YYY; in original form ZZZ}

\pubyear{2019}

\begin{document}
\label{firstpage}
\pagerange{\pageref{firstpage}--\pageref{lastpage}}
\maketitle

\begin{abstract}
Exploiting the sample of 30 local star-forming, undisturbed late-type galaxies in different environments drawn from the GAs Stripping Phenomena in galaxies with MUSE (GASP), we investigate the spatially resolved Star Formation Rate-Mass (\Ssfr-\Sm) relation. Our analysis includes also the galaxy outskirts (up to $>4$ effective radii, $r_e$), a regime poorly explored by other Integral Field Spectrograph surveys. Our observational strategy  allows us to detect \Ha out to more than 2.7$r_e$ for 75\% of the sample.  { Considering all galaxies together,} the correlation between the \Ssfr and \Sm is quite broad, with a scatter of 0.3 dex.  It gets steeper and shifts to higher \Sm values when external spaxels are excluded and moving from less to more massive galaxies. The broadness of the overall relation suggests galaxy-by-galaxy variations. Indeed, each object is characterized by a distinct \Ssfr-\Sm relation { and in some cases the correlation is very loose}. The scatter of the relation { mainly} arises from the existence of bright off-center star-forming knots whose \Ssfr-\Sm relation is systematically broader than that of the diffuse component.  
The \Ssfr-\Stg (total gas surface density) relation is as broad as the \Ssfr-\Sm relation, indicating that the surface gas density is not a primary driver of the relation.  
Even though a large galaxy-by-galaxy variation exists, mean  \Ssfr and \Sm values vary of at most 0.7 dex across galaxies. We investigate the relationship between the local and global SFR-M$_\ast$ relation, finding that the latter is driven by the existence of the size-mass relation.
\end{abstract}




\section{Introduction}

The Star Formation Rate (SFR) - Mass relation (\ma) is one of the most studied relations in astrophysics \citep[e.g.][]{Brinchmann2004, Salim2007, Noeske2007a, Noeske2007b}. It shows how tightly a galaxy's SFR and its total stellar mass are related, at any redshift, for star forming galaxies (Star Forming Main Sequence, SFMS).  Higher stellar mass systems undergo more intense star formation activity than lower mass systems. At all redshifts, the dispersion of the correlation is only 0.2-0.3 dex  in SFR for a fixed mass and the slope is somewhat smaller than unity, implying that the relative rate at which stars form in galaxies, i.e. the specific star formation rate (SSFR), declines weakly with increasing galaxy mass \citep{Salim2007, Schiminovich2007}. From   $z=0$  to high redshift the normalization  increases to  higher values: by $z\sim 2$ at a fixed stellar mass SFRs are higher by a factor of $\sim 20$  \citep[e.g.][]{Noeske2007a, Daddi2007, Vulcani2010, Elbaz2011, Whitaker2012, Speagle2014, Schreiber2015, Barro2017}.

The existence of such relation points to a scenario where galaxies form through  secular processes rather than stochastic merger-driven star-forming episodes. Universal laws seem to govern their evolution throughout cosmological time and across many different environments. Galaxies evolve along the SFMS, increasing in mass through the accretion of cold gas through mergers and/or from the cosmic web, until the supply of gas is shut off, most likely when a critical mass is reached \citep{Cattaneo2006}. Star formation is thus interrupted, and the galaxy moves to the red sequence, where it may further grow in mass and size through minor mergers \citep[e.g.,][]{Faber2007, Lilly2013}. The quenching phase is critical in the life of a galaxy, nonetheless, the conditions leading the the galaxy quenching are still  obscure. 

A great limitation to the comprehension of the connection between SFR and \ma is that most studies consider galaxies as a whole, with integrated observations that do not distinguish among the morphological components, or  that only partially cover the galaxies and are thus subject to aperture effects. A step forward is to characterise the relation on smaller scales, using spatially resolved data. 

If the relation between star formation and mass still held on small scales, it would suggest that the global SFR-\ma relation is primarily the outcome of the local correlation and the mechanism that drives the star formation activity with respect to the stellar mass could be universal across various physical scales, similarly  to the Kennicutt-Schmidt relation \citep{Kennicutt1998b} between the SFR surface density  (\Ssfr) and the cold gas surface density that has been found to hold both locally \citep[e.g.,][]{Bigiel2008} and globally. The opposite is indeed unreasonable, as it seems rather fine-tuned that a mechanism acting on global scales affects both the surface mass density (\Sm) and  \Ssfr 
 exactly in the same way, without a local process being involved or that a process reverts to the mean when averaged over large volumes.

On the other hand, the lack of a correlation between these spatially resolved quantities would suggest a galaxy-wide process that regulates the SFR of galaxies as a whole.

Outside our Galaxy, spatially resolved properties have been investigated for M31 \citep{Viaene2014}, who performed SED fitting of a panchromatic dataset, covering UV to submm wavelengths at a physical resolution of 140$\times$ 600 pc. They found that regions of higher \Sm  tend to host less star formation, suggesting an internal downsizing process in star formation, in which the regions of highest stellar density have already stopped forming stars. In addition, star formation at small scales has a weak dependence on the stellar mass: small- scale environments characterised by different stellar masses can be associated with very different levels of star formation.

An attempt to study the spatially resolved SFMS relation in local  (0.01 < z < 0.02) massive disc galaxies using photometric data was performed by \cite{Abdurrouf2017}, who combined seven bands 
imaging data from Galaxy Evolution Explorer (GALEX) and Sloan Digital Sky Survey (SDSS) 
fitting the spatially resolved SED of a galaxy with a set of model photometric SEDs using a Bayesian statistics approach. They showed that the relation is almost linear and presents a bending at \Sm$ \rm = 3.1\times  10^{7} \, M_\odot \, kpc^{-2}$. The same authors also applied the same technique at higher redshift, finding overall similar results \citep{Abdurrouf2018}. Exploiting 10-band photometry from the UV to the near-infrared at HST resolution, \cite{Morselli2018} derived spatially resolved maps of stellar masses and SFRs, finding that the star formation activity is centrally enhanced in galaxies above the SFMS and centrally suppressed below it. The level of suppression correlates with the distance from the SFMS.
They concluded that the presence of the bulge plays an important role: 
at a fixed stellar mass, bulge-dominated galaxies are preferentially located below the SFMS, while disk-like galaxies with low Sersic indexes are preferentially located on 
the upper envelope of the SFMS.  The suppression of star formation activity in the central region of galaxies below the SFMS points to an inside-out quenching scenario, as star formation is still ongoing in the outer regions.  
The importance of separating the contribution of bulges and disks has also been emphasized by \cite{Abramson2014}, who argued that  re-normalizing the SFR by disk stellar mass, the star formation efficiency does not depend anymore on galaxy mass for star-forming disks, 
In contrast, \cite{Willett2015} found no statistically significant difference in the SFR-\ma relations in a wide range of nearby morphological sub-types of star-forming disk galaxies, suggesting that these systems are strongly self-regulated.

The great step forward to the comprehension of the connection between SFR and \ma is  due to the advent of the Integral Field Spectroscopy (IFS) surveys, such as ATLAS3D \citep{Cappellari2011}\footnote{Note that because of its focus on early type galaxies, ATLAS3D essentially avoids star forming systems, it is not an ideal sample to study the SFMS \citep{McDermid2015}.}, CALIFA \citep{Sanchez2012, Husemann2013, GarciaBenito2015}, SAMI \citep{Bryant2015}, and MaNGA \citep{Bundy2015}, that have been designed to better understand the star formation and quenching in galaxies and that should help  disentangle the contributions of different parameters to the SFR-\ma relation, in the local universe. These surveys observed from $\sim100$ to 10000 galaxies, with a spatial coverage ranging from at most 1 to 3 effective radii ($r_e$). A complete comparison among the different surveys is given in Tab. 3 of \cite{Bundy2015}.

All the largest IFU surveys have found overall concordant results, that can be summarised as follows. 
In nearby galaxies, \Sm and \Ssfr for regions where star formation dominates the ionization 
constitute a tight correlation \citep{RosalesOrtega2012, Sanchez2013, Pan2018, Liu2018, Erroz2019}. \cite{Hall2018} have found that the slopes and zero-points of the different SFMS probed at different spatial resolution  scales ranging from 0.05 to 10 kpc remain essentially the same. Also the scatter at each resolution scale is relatively constant out to 0.5-1 kpc, dropping slightly beyond that range \citep[see also][]{ Kruijssen2014}. 
This correlation is invariant to changes in global properties, such as inclination, total stellar mass, local environment, H{\sc i} mass,  redshift (\citealt{CanoDiaz2016, Ellison2018, Hall2018}, but see \citealt{Medling2018} for a dependence of the SFR distribution on galaxy local number density).

\cite{CanoDiaz2016} exploiting CALIFA data and \cite{Hsieh2017} investigating MaNGA data  have also shown that the  slope and dispersion of this local SFMS are similar to those of the global SFR-\ma relation \citep[see also][]{Hall2018}. This result has also been shown to hold at higher redshift \citep{Wuyts2013, Magdis2016}. The remarkable agreement between the local and global SFR and \ma relations suggests that the global main sequence may originate from a more fundamental relation on small scales and that  the SFR is controlled by the amount of old stars locally \citep{Hsieh2017}.

The scatter of the \Ssfr - \Sm relation is strongly related to Hubble type \citep{GonzalezDelgado2016}, with the relation becoming less tight  when early type galaxies are considered in the sample. In fact, a subpopulation of non passive early type galaxies exists \citep{Sarzi2010, Medling2018} and these are preferentially located below the SFMS. According to \cite{Medling2018}, the presence of a bulge plays only a marginal role: some galaxies with bulges continue their star formation exactly as late-type spirals would. On the other hand, \cite{GonzalezDelgado2016, Maragkoudakis2017} found that galaxies with higher bulge fractions tend to exhibit lower SFR for their \Sm. In M31, \cite{Viaene2014} found that the bulk of stars formed at early epochs in the bulge, and the more recent star formation happens at larger galactocentric distances (i.e. $>3$ kpc).

The clear role of Hubble type in defining the offset around the overall \Ssfr - \Sm relation suggests that some global morphology-related property modulates the local \Ssfr, such as the gas content \citep{RobertsHaynes1994, Tacconi2013}.

In spite of the many works focused on the spatially resolved \Ssfr - \Sm relation, no study has focused on the relation separately for bright HII regions and  diffuse component. 
Indeed, the two components give different information about the gas-stellar cycle. HII regions are photoionised by massive young stars formed recently; the diffuse component (also known as warm ionized medium, WIM) is a warm ($\sim 10^4$ K), ionized, low density (0.1 cm$^{-3}$) gas  between HII regions typically with low ionization parameter that can extend 1 kpc or more above the disk plane \citep[and references therein]{Mathis2000, Haffner2009}. This gas component is most likely ionized by stars (O, B and hot evolved stars) in the disk whose Lyman continuum photons travel large path lengths 
\citep{Mathis1986, Mathis2000, Domgorgen1994, Sembach2000, Wood2004, Wood2010}. Understanding if the two components follow similar \Ssfr-\Sm relations is thought important to understand their physics. Recently, a lot of effort is being put to characterise the metallicity of this diffuse component \citep[e.g.][]{Yan2018, Kumari2019}. \cite{Erroz2019} have contrasted the metallicity of the HII region and the diffuse component, finding that the former have on average  metallicities 0.1 dex higher than the latter.

In this paper we present the analysis of the spatially resolved SFR-\ma relation for 30 undisturbed late-type galaxies in different environments drawn from the GAs Stripping Phenomena in galaxies with MUSE (GASP\footnote{\url{http://web.oapd.inaf.it/gasp/index.html}}), 
 an  ESO Large programme that exploits the integral-field spectrograph MUSE mounted at the VLT with the aim to characterise where, how and why gas can get removed from galaxies in different environments. A complete description of the survey strategy, data reduction and analysis procedures is presented in \citet[][Paper I]{Poggianti2017a}. 

Sections \ref{sec:data}  presents the data sample and the data analysis, while Section  \ref{sec:results} includes the results. We  study the spatially resolved SFR-\ma relation (\Ssfr-\Sm) for all galaxies (\S \ref{sec:sfr_all}) and also for each galaxy separately  (\S \ref{sec:sfr_gal}), emphasizing the galaxy-by-galaxy variations. We  show how the variation is mainly due to the presence of bright off-center \Ha knots, whose spatial distribution is different from one object to the other and whose \Ssfr is not strictly related to the value of \Sm in the same resolution element. 
We also investigate the relationship between \Ssfr and total gas surface density (\Stg) (\S \ref{sec:gas}) and connect the local and global relations (\S \ref{sec:global}), showing how the latter is a byproduct of the galaxy size mass relation.  In Section \ref{sec:disc} we  discuss the results and conclude.

Throughout all the papers of the GASP series, we adopt a \cite{Chabrier2003} initial mass function (IMF) in the mass range 0.1-100 M$_{\odot}$. The cosmological constants assumed are $\Omega_m=0.3$, $\Omega_{\Lambda}=0.7$ and H$_0=70$ km s$^{-1}$ Mpc$^{-1}$.

\section{Data sample, observations and  data analysis} \label{sec:data}
The GASP targets are at redshift $0.04<z<0.1$, span a wide range of galaxy stellar masses, from $10^9$
to $10^{11.5} M_{\odot}$ and are located in
different environments (galaxy clusters, groups, filaments and
isolated).  The GASP project observed 114 galaxies in total. 76 of them are in clusters and 38 of them are in the less massive environments.  The sample includes both galaxies selected as stripping candidates and undisturbed galaxies.
A complete description of the survey strategy, observations, data reduction and analysis procedure is presented in \citetalias{Poggianti2017a}. 

We stress  that our observations cover the entire optical extension of the galaxy (see below for the definition of the galaxy sizes), up to several effective radii, so our data are not affected by aperture loss.  

\begin{figure}
\centering
\includegraphics[scale=0.35]{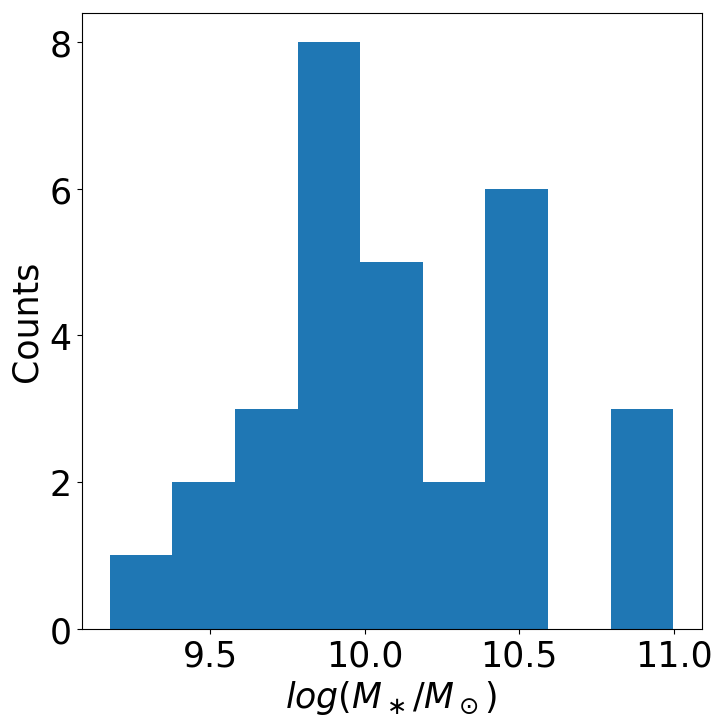}
\caption{Total stellar mass distribution of the galaxies analysed in this work.\label{fig:mass} }
\end{figure}

\begin{figure*}
\centering
\includegraphics[scale=0.35]{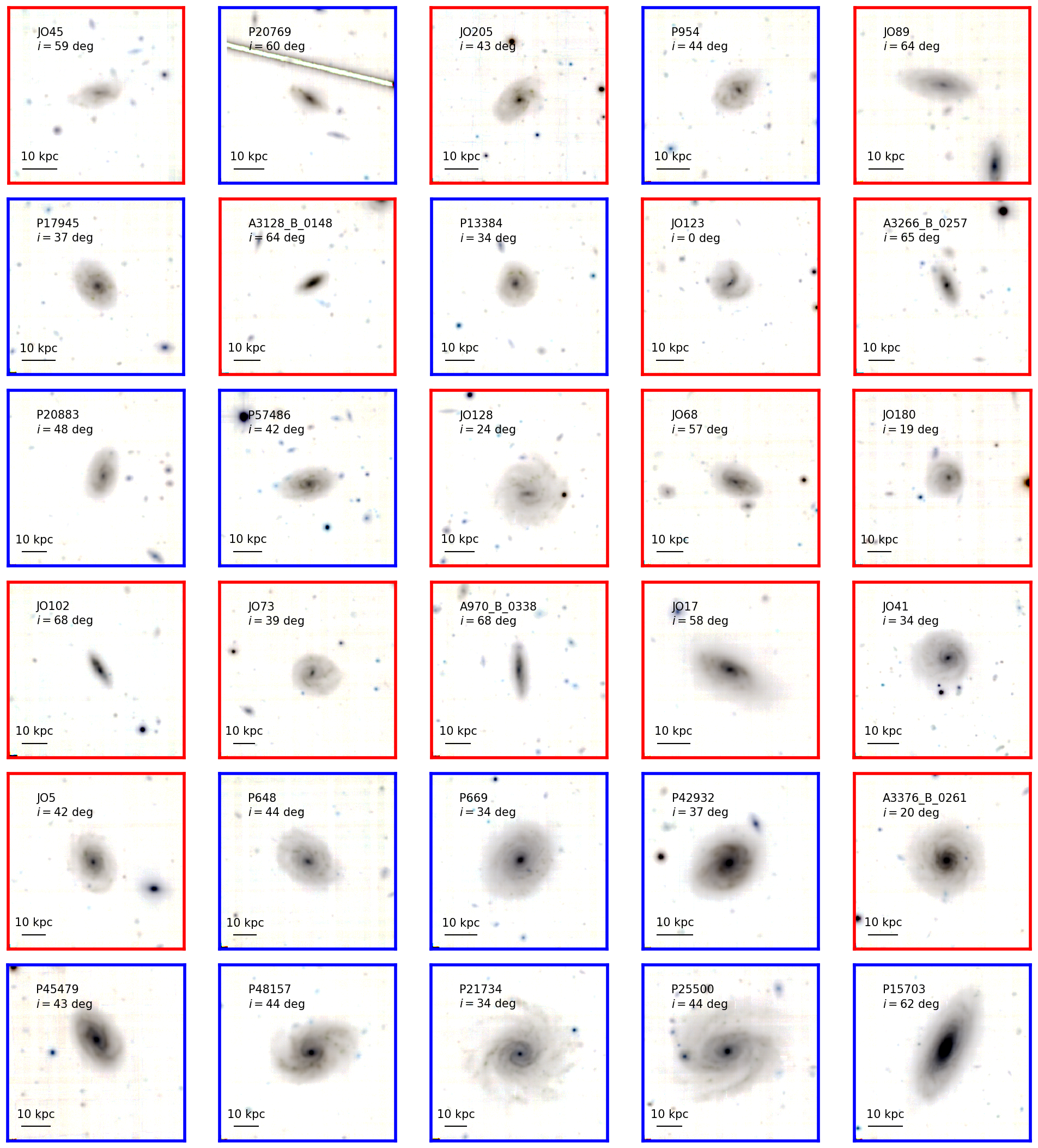}
\caption{RGB images of the galaxies used in this paper, sorted by increasing stellar mass. The reconstructed $g$, $r$, $i$  filters from the MUSE cube have been used. Colour map is inverted for display purposes.
North is up, and east is left. Galaxies surrounded by a red square belong to clusters, galaxies surrounded by a blue square belong to the field. In each panel, the galaxy inclination is also given.
\label{fig:color_images} }
\end{figure*}

\subsection{Data sample}
For this work we exploit the control sample selected in \citet[Paper XIV]{Vulcani2018c}. They extracted 
from the cluster+field control sample those galaxies that indeed are undisturbed and do not show any clear  sign of environmental effects (ram pressure stripping, tidal interaction, mergers, gas accretion...) on their spatially resolved star formation distribution, for a total of 16 cluster members and 14 field galaxies. 
We refer to Table 2 of \citetalias{Vulcani2018c} for the list of the objects, along with redshifts, coordinates, integrated stellar masses and star formation rates.  We remove from their sample JO93 and P19482 that, after a careful inspection of their \Ha maps, turned out to be in an initial phase of stripping. Figure \ref{fig:mass} shows the total stellar mass distribution of the galaxies entering the sample, while Fig. \ref{fig:color_images} presents an overview of the sample, sorted by increasing total stellar mass,  and shows  the colour composite images of the targets, obtained combining the reconstructed $g-$, $r-$ and $i-$filters from the MUSE datacube. As galaxies are drawn from different environments, we highlight with a red square the galaxies belonging to the cluster sample, with a blue square those belonging to the field. This colour scheme will be adopted throughout the paper.  Cluster galaxies are found at intermediate projected cluster centric distances, in the range 0.7-1.3 R$_{200}$,

\subsection{Data analysis}\label{sec:analysis}

As extensively presented in \citetalias{Poggianti2017a}, we corrected the reduced datacube for extinction due to our Galaxy and subtracted the stellar-only component of each spectrum
derived with our spectrophotometric code \sinopsis \citep{Fritz2017}.  
In addition to the best fit stellar-only model cube that is subtracted from the observed cube, \sinopsis provides for each MUSE spaxel stellar masses, luminosity-weighted and mass-weighted ages and star formation histories in four broad age bins.

We then derived emission line fluxes with associated errors using \kube \citep{Fossati2016}, an IDL public software.  We consider as reliable only spaxels with S/N(\Ha)>5.
\Ha luminosities corrected both for stellar absorption and for dust
extinction were used to compute SFRs, adopting the \cite{Kennicutt1998a}'s relation: $\rm SFR (M_{\odot}
\, yr^{-1}) = 4.6 \times 10^{-42} L_{\rm H\alpha} (erg \, s^{-2})$. 
The extinction is estimated from the Balmer decrement
assuming a value $\rm H\alpha/H\beta = 2.86$ and the \cite{Cardelli1989} extinction law. 
The MUSE data reach a surface brightness detection limit of $\rm V\sim 27 \, mag \, arcsec^{-2}$ and $\rm  H\alpha \sim 10^{-17.6} \, erg \, s^{-1}\,  cm^{-2} \,  arcsec^{-2}$ at the 3$\sigma$ confidence level \citepalias{Poggianti2017a}, which translate into a \Ssfr limit of $\rm \sim 7 \times 10^{-5} \, M_\odot \, yr^{-1} \, kpc^{-2}$.

We used the extinction also to derive the gas mass density,  using the empirical relation 
discussed in 
 \cite{Barrera2018}: $\rm \Sigma_{tot \, gas}=30 \times \left( A_v/mag \right) \, [M_\odot \, pc^{-2}]$. This relationship was calibrated using the observed gas mass density from spatially resolved observations of CO and the V-band attenuation for galaxies in the CALIFA EDGE survey \citep{Bolatto2017}.

We employed the standard diagnostic diagram
[OIII]5007/$\rm H\beta$ vs [OI]6300/$\rm H\alpha$ to separate the regions powered by star formation from regions powered by 
AGN or LINER emission. Only spaxels with an S/N$>$3 in all emission lines involved are considered. 
We adopted the division lines by \citet{Kauffmann2003b}. 
For the majority of the galaxies most of the \Ha is powered by photoionization (plots not shown),
and no galaxy in the sample hosts an AGN in its center. 
To compute SFRs, we considered only the spaxels whose ionised flux is powered by Star Formation. 

As largely discussed by \citet[Paper XIII]{Poggianti2019}, the diagnostic diagram based on the  [OI] is generally  very sensitive to physical processes different from Star Formation, such as thermal conduction from the surrounding hot ICM, turbulence and shocks. Our choice to use the [OI]-based diagnostic diagram instead of other available line-ratio diagrams
can therefore be considered as a conservative lower limit of the real star formation budget. 
Appendix \ref{app:NII} will discuss some results using the [NII]-based diagnostic diagram, for comparison. 


For each spaxel in each galaxy we also computed the galactocentric radius fixing the centre of the galaxy
to the peak of the stellar mass map. The radius is then expressed in units of  $r_e$, which is computed on I-band images by measuring the radius of an ellipse including half of the total light of the galaxy (Vulcani et al. 2019a - Paper XVI - and Franchetto et al. in prep.). We remind the reader that our observational strategy allows us to detect \Ha also in the galaxy outskirts: 75\% of the sample have a \Ha coverage larger than 2.7$r_e$. This maximum extension of the \Ha disk in units of $r_e$, 
is independent of stellar mass (plot not shown). 

We correct all quantities for the effect of inclination. We note however that no edge-on galaxy is present in the sample: galaxies span the range $20\degree < i < 70\degree$.

\section{Results}\label{sec:results}

\subsection{The \Ssfr-\Sm relation of all galaxies }\label{sec:sfr_all}
The total integrated SFR-\ma relation for the sample has been presented in \citetalias{Vulcani2018c},\footnote{Note that in \citetalias{Vulcani2018c} we identified regions powered by star formation using the diagnostic diagrams based on the [N{\sc II}] line rather than those based on the [O{\sc I}] line. However, this choice does not affect the results, therefore we do not discuss again the global SFR-\ma relation of the sample.} here we focus on the spatially resolved relation. 

\begin{figure*}
\centering
\includegraphics[scale=0.55]{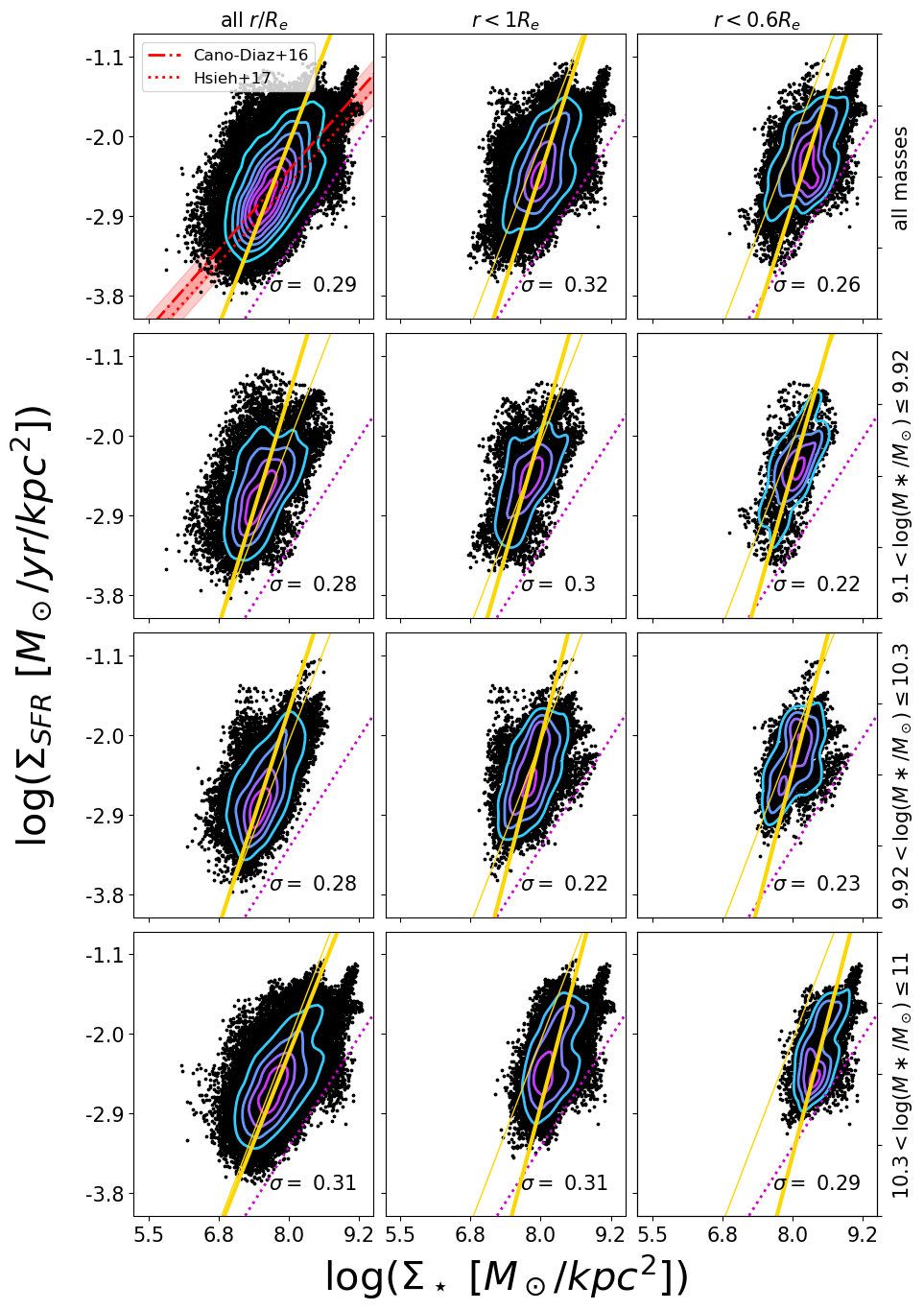}
\caption{Spatially resolved SFR-\ma relation using all spaxels of all galaxies in the sample (left), the spaxels within 1$r_e$ (central) and the spaxels within 0.6$r_e$ (right), both for all galaxies together (upper row) and for galaxies in three different mass bins, as indicated in the left labels. Thick gold lines show the fit to the relation, the thin gold line shows the fit obtained in the first panel, for comparison. In the first panel, results from \citet{CanoDiaz2016, Hsieh2017} are reported in red, for comparison, after converted to the same IMF. Contours identify regions characterised of similar density of points. { The magenta dotted line represents the effective threshold in spatially resolved specific SFR entailed by the adopted cuts in S/N corresponds to $\rm 10^{-11.2}\, yr^{-1} \, kpc^{-2}$.} 
\label{fig:sfr_mass_all} } 
\end{figure*}

The upper left panel of Figure \ref{fig:sfr_mass_all} shows the \Ssfr-\Sm relation, using all the 92020 star forming spaxels included in our sample. { The effective threshold in spatially resolved specific SFR entailed by the adopted cuts in S/N corresponds to $\rm 10^{-11.2}\, yr^{-1} \, kpc^{-2}$ and it is also shown. By selection, 99\% of the spaxels stay above the threshold.} A correlation between the two quantities is immediately visible, with more massive spaxels typically having higher values of SFR, even though the scatter of the relation is quite large. { To compute the scatter, we subdivided the sample in 10 \Sm bins and computed the standard deviation of \Ssfr in each bin separately. We then took the mean value as typical scatter, which turns out to be } 
$\sim0.3$ dex.  The correlation spans almost five orders of magnitude in \Sm and three in \Ssfr.

We compare our results with those presented in \cite{CanoDiaz2016, Hsieh2017}. Both studies have  highlighted that the \Ssfr-\Sm relation is rather tight, having a standard deviation of at most  0.2 dex, and has a slope of $\sim 0.7$. \cite{CanoDiaz2016}  based their analysis on CALIFA data, and analysed 306 galaxies of all morphological types  and environments, with inclinations $< 60\degree$,  in the redshift range $0.005 < z < 0.03$ and in the mass range  $10^{9.7} <$ \ms $ < 10^{11.4}$. The CALIFA observational strategy guarantees a coverage of the spatial extension of the galaxies up to 2.5 effective radii, and spatial resolution of $\sim$1kpc at the average redshift of the survey. To select regions powered by star formation, they used both the BPT  diagrams based on  [N II] and the WHAN diagram \citep{CidFernandes2011} (EW(\Ha) $>6 \AA{}$).

The analysis by \cite{Hsieh2017} is based on MaNGA data. Their targets were selected to represent the overall galaxy population with stellar masses greater than $10^9$\ms at $0.01 < z < 0.15$.  1085 non interacting galaxies entered their selection. The MaNGA observational strategy guarantees a coverage of the spatial extension of the galaxies up to 1-5-2.5 effective radii, and spatial resolution of $\sim$1kpc at the average redshift of the survey. 
To select regions powered by star formation, they used both the BPT  diagrams based on  [S II] and a modified version of the WHAN diagram \citep{CidFernandes2011} ($\log([N II]/$\Ha)<-0.4 and EW(\Ha)$ > 5\AA{}$). To compute the fit to the relation, they applied the both the ordinary least squares method and the orthogonal distance regression.

Qualitatively, our relation, fitted with a linear least-squares regression, is much broader and steeper, having a slope of 1.56$\pm$0.07.
 Quantitative comparisons between ours and their analyses is not possible, as works did not specify the spatial extension coverage of the galaxies that entered their samples, nor the minimum \Sm and \Ssfr reached by their galaxies. In addition, different selection criteria have been adopted to select regions photoionised by star formation.  All these aspects, along with the sample selection, analyzing method, fitting recipe and spatial resolution,  play important roles in the determination of the slope of the fit. Overall, literature results range from 0.6 to $>1$ \citep[see also][]{Sanchez2013, Wuyts2013, Abdurrouf2017, Hsieh2017, Maragkoudakis2017}. 
 
To mimic the effect of the different spatial coverage on galaxies and different mass ranges probed, Fig.  \ref{fig:sfr_mass_all} also shows the  \Ssfr-\Sm relation using only the spaxels within 1 and 0.6 $r_e$, and in three different mass bins. It unveils that relations get steeper and shift to higher \Sm values when  spaxels at large galactocentric distances are excluded from the fit, as shown by the position of the fitting line in Fig.  \ref{fig:sfr_mass_all}. This is true  both for the entire population and for galaxies in a given stellar mass bin. At each given galactocentric distance, the fit also shifts to higher \Sm values going from low to high mass galaxies. It appears evident that relations have smaller scatter when considering only central spaxels and less massive galaxies, even though the smallest scatter is always $>0.2$ dex. Most of the spaxels with \Sm $\rm \leq 10^7 M_\odot \, pc^{-2}$ are found only in the external regions. 
Note that this is approximately the same radial regime  where \cite{Erroz2019} found a break down of the relation, above which the relation is steeper and below which it is flatter. 

Galaxy outskirts - most likely missed by previous larger surveys - might be subject to different physical processes and therefore follow less tight relations, broadening the overall trends. 

It is also interesting to note that in the most massive bin there are hints for the coexistence of two parallel relations: one produced by the spaxels in the  regions $<$0.6 $r_e$ and, shifted toward lower \Sm values, one produced by the spaxels in the intermediate regions ($0.6<r/r_e<1$). This behaviour is not visible at lower masses. The highest mass bin includes galaxies more massive than $ \rm 3\times 10^{10} M_\odot$, which is a well known discriminating value in galaxy evolution. \cite{Kauffmann2003a} were the first to show that galaxies less massive than this value have low surface mass densities, low concentration indices typical of discs and young stellar populations. More massive galaxies have high surface mass densities, high concentration indices typical of bulges, and predominantly old stellar populations.

\begin{figure*}
\centering
\includegraphics[scale=0.35]{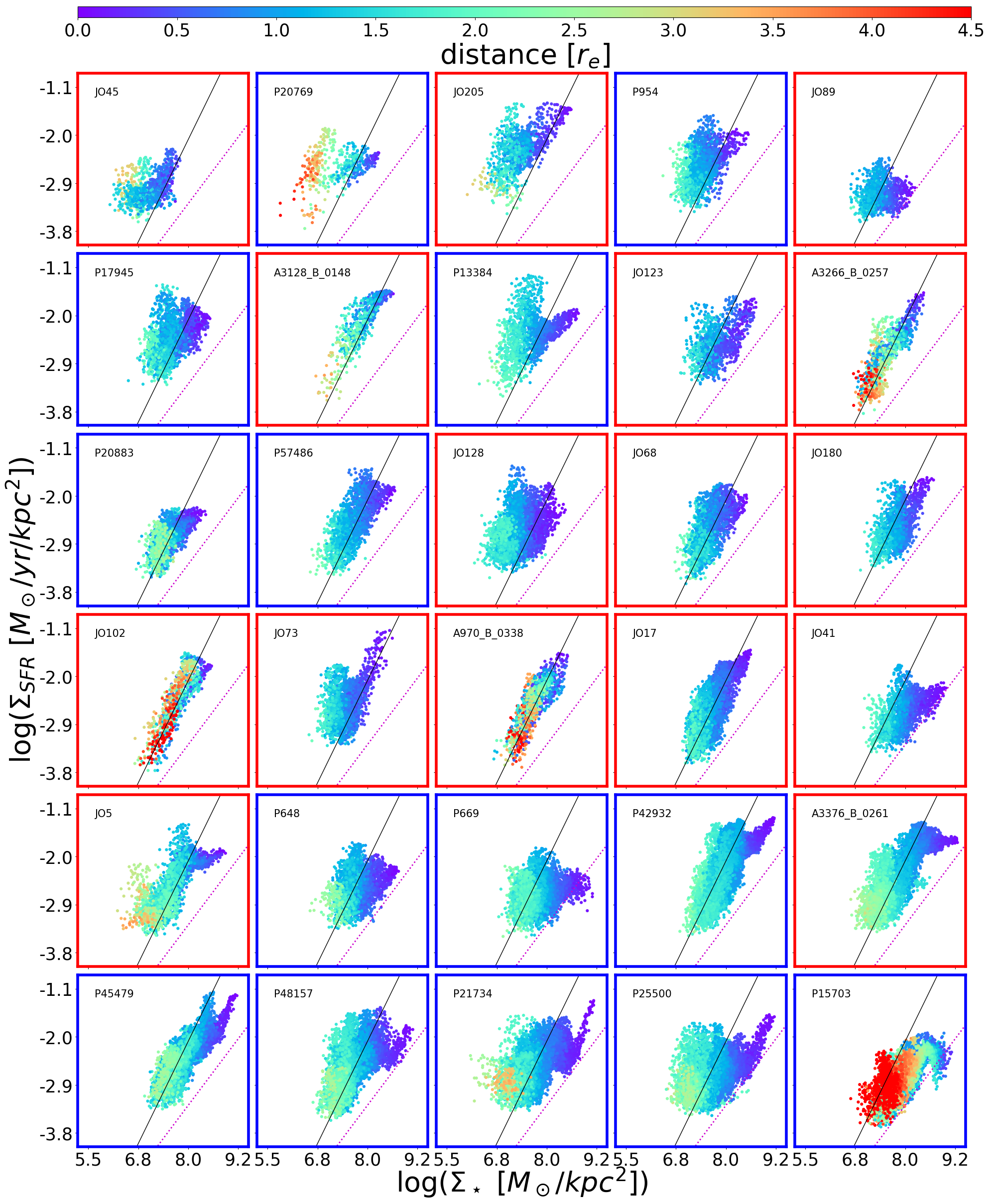}
\caption{\Ssfr - \Sm relation for all galaxies in the sample, sorted by increasing stellar mass and colour coded by the galactocentric distance of each spaxel, in unit of $r_e$.  Galaxies surrounded by a red square belong to clusters, galaxies surrounded by a blue square belong to the field. The black line represents the fit to the whole sample, from Fig.\ref{fig:sfr_mass_all}. { The magenta dotted line represents the effective threshold in spatially resolved specific SFR entailed by the adopted cuts in S/N corresponds to $\rm 10^{-11.2}\, yr^{-1} \, kpc^{-2}$.} \label{fig:sfr_mass_dist_mass} }
\end{figure*}

We note that, plotting at the same time all galaxies together, it is not clear whether they all follow similar relations or if each galaxy is characterized by a different slope, intercept, scatter. In the next subsection we therefore characterise each galaxy separately.

\subsection{The galaxy-by-galaxy  \Ssfr-\Sm relation }\label{sec:sfr_gal}

Figure \ref{fig:sfr_mass_dist_mass}  presents the \Ssfr-\Sm relation for each galaxy separately, distinguishing among spaxels at different galactocentric distances. It appears evident that, even though overall { in most cases} a correlation does exist,  each object spans a distinct locus on the \Ssfr-\Sm plane \citep[in agreement with ][]{Hall2018} and the relation that better describes it  is characterized by a different slope and scatter. { Table \ref{tab:values} presents the slope, intercept, coefficient correlation value (r\_value), the scatter and the standard error of the estimate for each galaxy separately. Half of the sample has r\_value$<0.5$, indicating there is no correlation between the \Ssfr and \Sm and  suggesting that fitting relations is meaningless. Only 6/30 galaxies have  r\_value$>0.8$, suggesting a rather strong correlation. In all cases the p-value is close to zero, supporting the reliability of the results. }

{The slope of the relation for each of the individual galaxies is close to one, especially for those with the highest r\_value, and even shallower when the correlation is weaker. The steep slope of the global relation (Fig.~\ref{fig:sfr_mass_all}) is therefore probably a consequence of the different intercepts of the galaxies in the sample.}

\begin{table}
\caption{Least-squares regression parameters for each galaxy.   \label{tab:values}}
\centering
\begin{tabular}{|l|r|r|r|r|r|}
\hline
  \multicolumn{1}{|c|}{ID} &
  \multicolumn{1}{c|}{slope} &
  \multicolumn{1}{c|}{intercept} &
  \multicolumn{1}{c|}{r\_value} &
  \multicolumn{1}{c|}{std\_err} &
  \multicolumn{1}{c|}{scatter} \\
\hline
  A3128\_B\_0148 & 1.01 & -10.04 & 0.866 & 0.024 & 0.24\\
  JO102 & 1.2 & -11.67 & 0.852 & 0.025 & 0.27\\
  A3266\_B\_0257 & 1.05 & -10.6 & 0.833 & 0.023 & 0.28\\
  P45479 & 0.8 & -8.65 & 0.82 & 0.008 & 0.19\\
  A3376\_B\_0261 & 0.84 & -8.91 & 0.817 & 0.007 & 0.24\\
  A970\_B\_0338 & 1.06 & -10.62 & 0.815 & 0.023 & 0.23\\
  P15703 & 0.63 & -7.72 & 0.765 & 0.007 & 0.22\\
  JO5 & 0.76 & -8.25 & 0.763 & 0.012 & 0.21\\
  JO17 & 1.08 & -10.9 & 0.761 & 0.015 & 0.25\\
  P42932 & 0.91 & -9.47 & 0.745 & 0.009 & 0.23\\
  P57486 & 0.88 & -9.18 & 0.718 & 0.018 & 0.2\\
  P20883 & 0.62 & -7.43 & 0.653 & 0.016 & 0.22\\
  JO205 & 0.56 & -6.37 & 0.626 & 0.021 & 0.23\\
  JO68 & 0.8 & -8.67 & 0.617 & 0.023 & 0.29\\
  P21734 & 0.48 & -6.21 & 0.605 & 0.007 & 0.22\\
  P48157 & 0.61 & -7.12 & 0.595 & 0.01 & 0.26\\
  JO180 & 0.66 & -7.7 & 0.533 & 0.026 & 0.16\\
  JO41 & 0.44 & -6.13 & 0.525 & 0.016 & 0.21\\
  JO123 & 0.47 & -6.18 & 0.502 & 0.024 & 0.22\\
  JO73 & 0.57 & -6.77 & 0.471 & 0.024 & 0.33\\
  P954 & 0.51 & -6.33 & 0.466 & 0.022 & 0.28\\
  JO45 & 0.33 & -5.25 & 0.442 & 0.022 & 0.17\\
  P17945 & 0.45 & -5.82 & 0.435 & 0.019 & 0.28\\
  P25500 & 0.3 & -4.94 & 0.41 & 0.007 & 0.27\\
  P13384 & 0.34 & -4.93 & 0.378 & 0.021 & 0.25\\
  P648 & 0.33 & -5.11 & 0.347 & 0.015 & 0.19\\
  JO128 & 0.34 & -5.12 & 0.343 & 0.014 & 0.28\\
  P20769 & 0.18 & -3.9 & 0.323 & 0.024 & 0.21\\
  P669 & 0.15 & -3.83 & 0.22 & 0.009 & 0.23\\
  JO89 & 0.07 & -3.57 & 0.09 & 0.022 & 0.19\\
\hline\end{tabular}
\end{table}

We note that masking the spaxels most likely located in the galaxy  bulge -  whose size has been obtained applying a 2D fitting on the i-band images (see Franchetto et al. in prep.)-  does not affect the results. We therefore include them in the forthcoming analysis. 

Some galaxies show quite elongated sequences (e.g. A3128\_B\_0148, P57486), while in some other cases a cloud rather than a sequence is observed (e.g. JO45, JO89). Few galaxies really show a flattening (A3376\_B\_0261, P669) or bending (P48157, P15703) of the relation at higher \Sm values. Note that this behaviour is not due to the presence of the bulge, as trends stay also when the contribution of the bulge is removed (plot not shown). Moretti et al. (in prep.) will provide a detailed description of these objects presenting anomalies in the central regions.

For some galaxies the scatter increases with decreasing \Sm (e.g. P25500), for some others different sequences coexist (e.g. P20769, P13384, JO5). Only a few galaxies show a really tight and unique sequence (e.g. A3128\_B\_0148, JO102, A970\_B\_0338). The latter are all in clusters. Apart from that, no other outstanding environmental effects are evident. This result is somehow expected, given that the cluster members in the control  sample most likely just entered the cluster environment from the field and have had not time yet to feel cluster specific processes. This is confirmed by the location in projected phase-space (plot not shown): their cluster centric distribution spans from 0.7 to 1.3 $R_{200}$, and their relative velocities are 
within $|\Delta(v)|/\sigma_{cl}$<2, where $\Delta(v)$ is the galaxy velocity with respect to the velocity of the cluster and $\sigma_{cl}$ is the velocity dispersion of the cluster, consistent with the fact that they have not approached yet the cluster center.

\begin{figure*}
\centering
\includegraphics[scale=0.35]{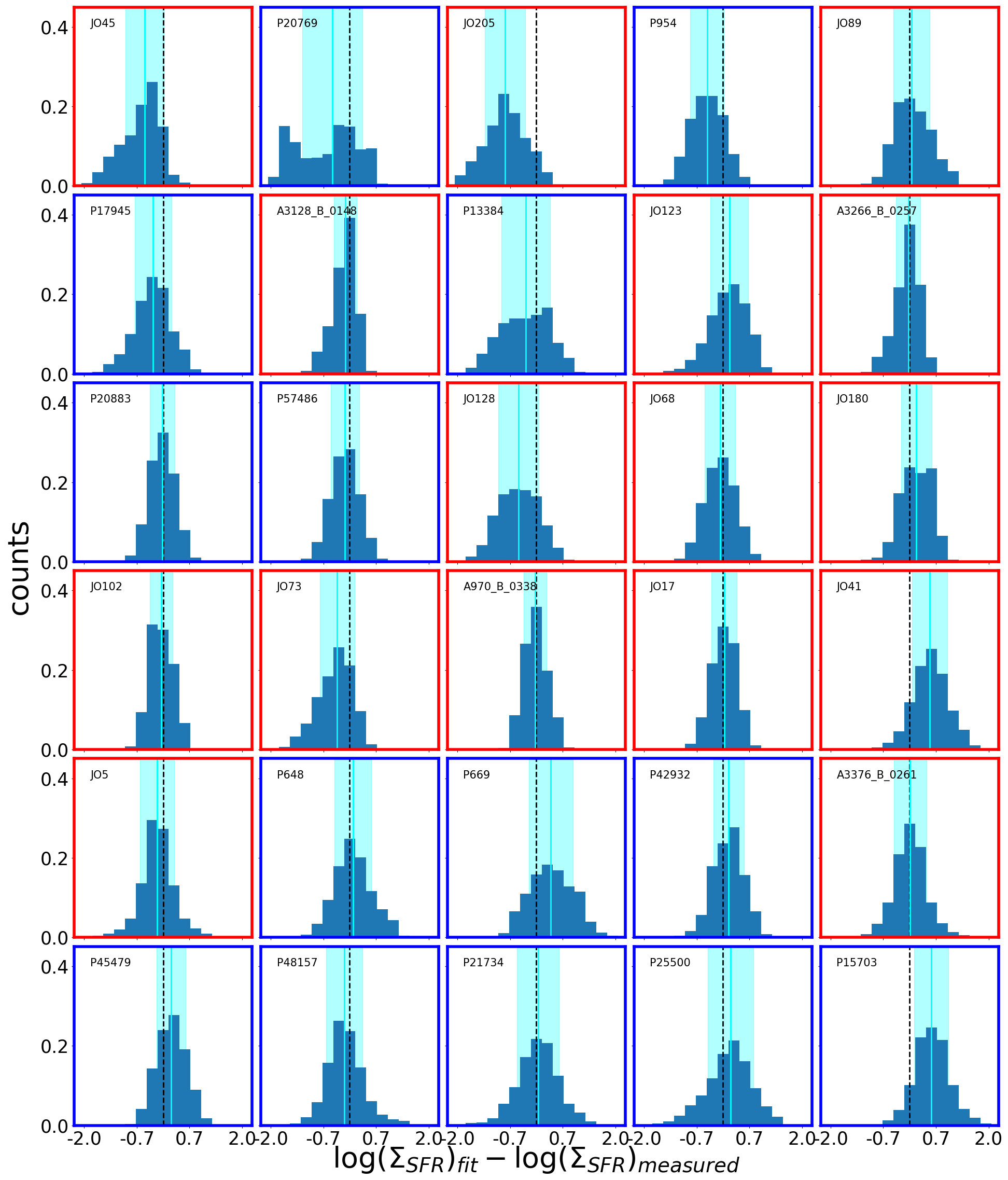}
\caption{{ For each galaxy, distribution of the differences between the galaxy \Ssfr and their expected value according to the fit to the entire sample (Fig.\ref{fig:sfr_mass_all}), given their \Sm. The black line is always centered at 0, the cyan lines and shaded area give median values and the standard deviation of the distribution ($1\sigma$). Galaxies are sorted by increasing stellar mass. Galaxies surrounded by a red square belong to clusters, galaxies surrounded by a blue square belong to the field.} \label{fig:sfr_delta_distr_mass} }
\end{figure*}

\begin{figure*}
\centering
\includegraphics[scale=0.33]{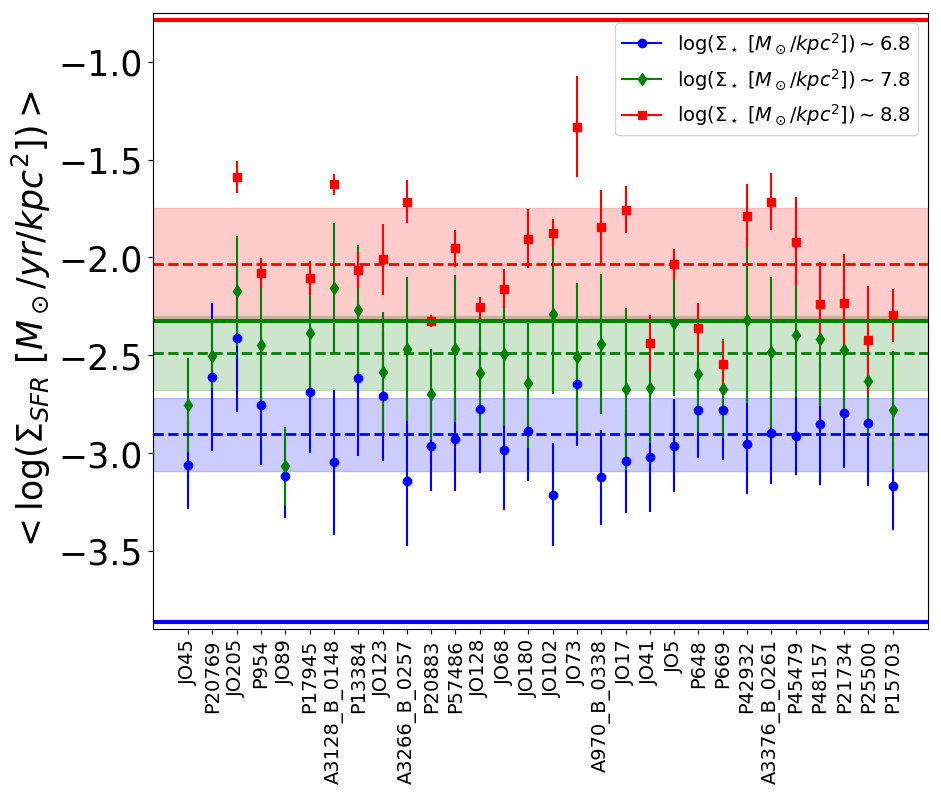}
\includegraphics[scale=0.33]{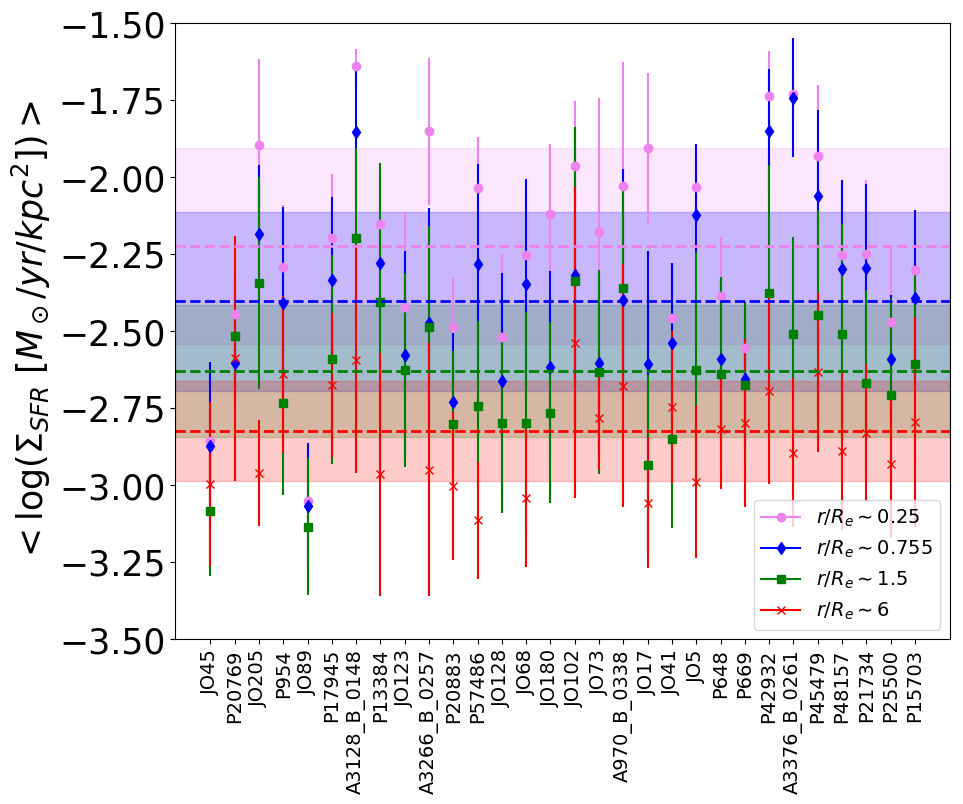}
\caption{Median \Ssfr in three \Sm bins (left) and in four galactocentric distance bins (right), for all galaxies in the sample, sorted by increasing stellar mass. In both panels error bars represent the standard deviation. { In the left panel, dashed horizontal lines represent median values in each \Sm bin, solid lines represent expected values given the fit of the total relation for that bin. In the right panel, dashed horizontal lines represent median values in each  bin of distance}. Shaded areas are the standard deviation of the medians.
\label{fig:sfr_mass_median_mass} }
\end{figure*}

{ To better highlight the galaxy- by-galaxy variations with respect to the relation obtained considering all galaxies together, for each galaxy we compute the difference between the galaxy \Ssfr and their expected value according to the fit to the entire sample (Fig.\ref{fig:sfr_mass_all}), given their \Sm. Figure \ref{fig:sfr_delta_distr_mass} shows the distribution of these differences for each galaxy separately. If a galaxy had the same \Ssfr-\Sm as the entire sample, the median value of the distribution would be 0 (marked by a black line in the plot). In each panel, the cyan line shows the median of the distribution and in many cases it coincides with the black line (e.g. JO17, JO68, A3376\_B\_0261). In contrast, for some galaxies (e.g. JO45, JO41, JO205, P15703), the median of the distribution is significantly different, suggesting that the  \Ssfr-\Sm of the galaxy  significantly deviates from the total one. In addition to the median values, also the shapes of the distributions give important information: while in some cases they are quite narrow (e.g. A970\_B\_0338, A3128\_B\_0148), in many other cases they are broad (e.g. P13384), show asymmetric tails (e.g. JO45) and are even double peaked (e.g. P20769). These results highlight that no a universal \Ssfr-\Sm relation exists for the galaxies.}

Figure \ref{fig:sfr_mass_dist_mass} also allows to inspect the position of each spaxel with respect to the galaxy center. Considering galactocentric distance, we find that
 spaxels in the external regions are characterised by  systematic lower \Sm values, and spaxels at similar distance  in terms of $r_e$ are distributed in the \Ssfr-\Sm plot on almost vertical relations, suggesting that the correlation between 
\Ssfr and distance might be stronger than \Ssfr and \Sm. Nonetheless,  if we contrast
 the \Ssfr  and distance, we find relations as broad as those presented in  Fig. \ref{fig:sfr_mass_dist_mass}, having a similar scatter (plot not shown). 

To better characterise the galaxy-by-galaxy variations, we consider separately the spaxels in three equally spaced mass density bins, centered at \Sm $\sim 6.3\times 10^{6}$ \msk,  $\sim 6.3\times 10^{7}$ \msk, $\sim 6.3\times 10^{8}$ \msk, and compute the median \Ssfr along with the scatter, measured as one standard deviation. The left panel of Fig. \ref{fig:sfr_mass_median_mass} shows these values for each galaxy separately, sorted by increasing total stellar mass. The scatter 
ranges from 0.2 to 0.3 dex in the three bins.  Even though almost all the plotted points agree within the errors with the median values, this Figure shows again that each galaxy is characterized by a different slope and scatter in the \Ssfr-\Sm relation. { In addition, it appears evident that while at intermediate values of \Sm  galaxies have \Ssfr values similar to what expected from the total fit, both at low and high \Sm values they strongly deviate from it, being systematically higher and lower, respectively.  }

 The right panel of Fig. \ref{fig:sfr_mass_median_mass} shows the median \Ssfr along with the scatter of the spaxels in four different galactocentric distance bins. The galaxy by galaxy variation is outstanding, especially for r/r$_e <$1, suggesting that regions characterised by high values of star formation can be present at different distances in different galaxies.  The scatter is always large, ranging from 0.15 to 0.35 dex. In the next section we will focus on these off-center star forming regions.

{ Taken together, Figures \ref{fig:sfr_mass_dist_mass} and \ref{fig:sfr_mass_median_mass} indicate that variations are evident not only among different galaxies, but also among regions within galaxies (defined according to galactocentric distance, \Ssfr or \Sm), implying that there is no ``universal" relation between \Sm  and \Ssfr.}

\begin{figure*}
\centering
\includegraphics[scale=0.35]{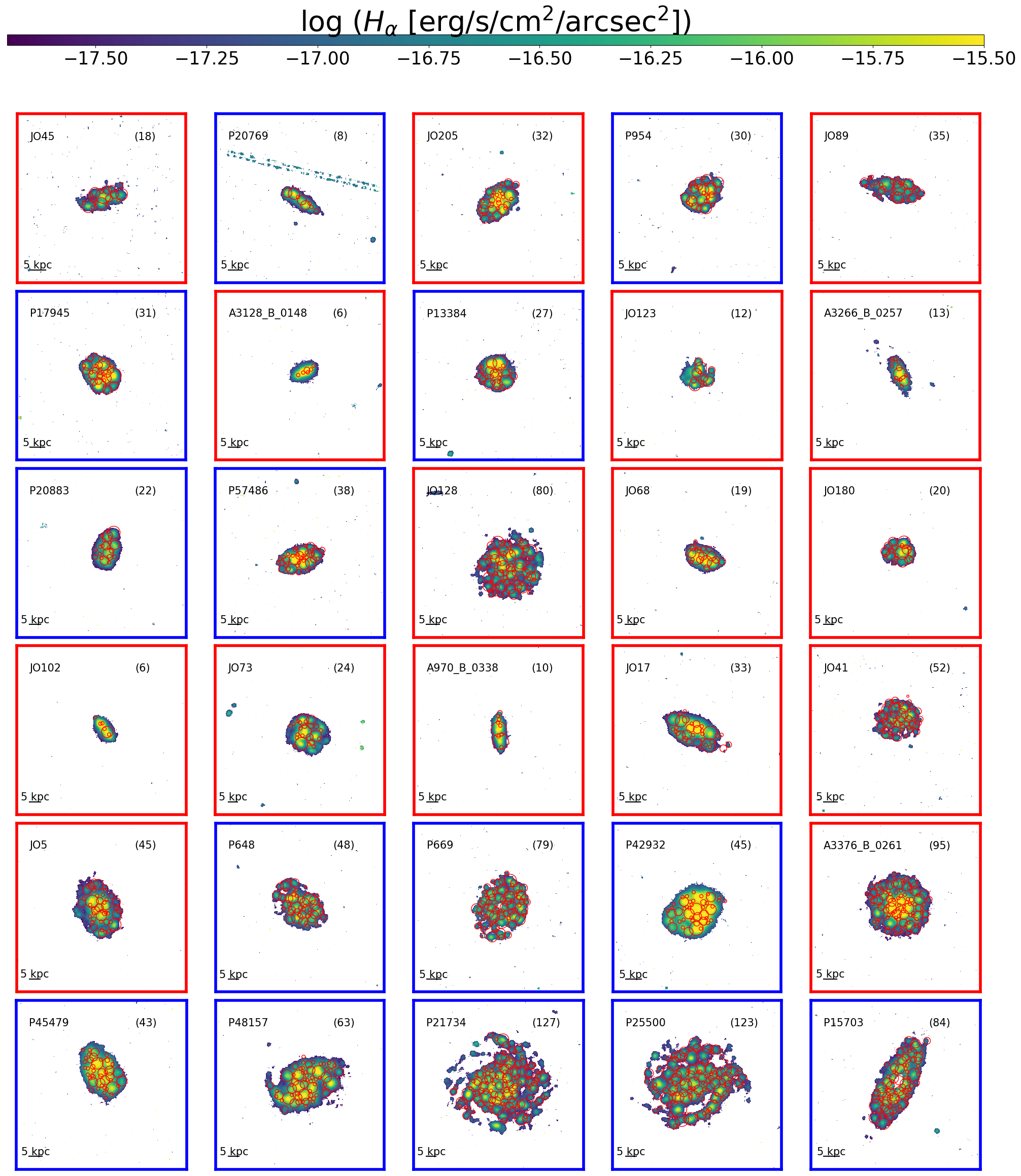}
\caption{\Ha maps for all galaxies in the sample, sorted by increasing stellar mass. Red circles show the \Ha knots.  Galaxies surrounded by a red square belong to clusters, galaxies surrounded by a blue square belong to the field. Numbers in parenthesis are the number of knots in each galaxy. \label{fig:Hamaps} }
\end{figure*}

We note that we have investigated whether the magnitude of the scatter depends on a number of global parameters, namely the total stellar mass, the SFR,  the $\rm \Delta (SFR)$ - i.e. the differences between the galaxy SFRs and their expected value according to the fit of the SFR-\ma relation, given their mass \citepalias{Vulcani2018c}, the galaxy inclination, the number of knots, the SFR in the knots, the SFR in the knots divided by the total stellar mass, the effective radius, but found no statistically significant trends. Appendix \ref{app:scatter} better discusses this point.

\subsubsection{Star forming knots and diffuse emission}
\begin{figure*}
\centering
\includegraphics[scale=0.35]{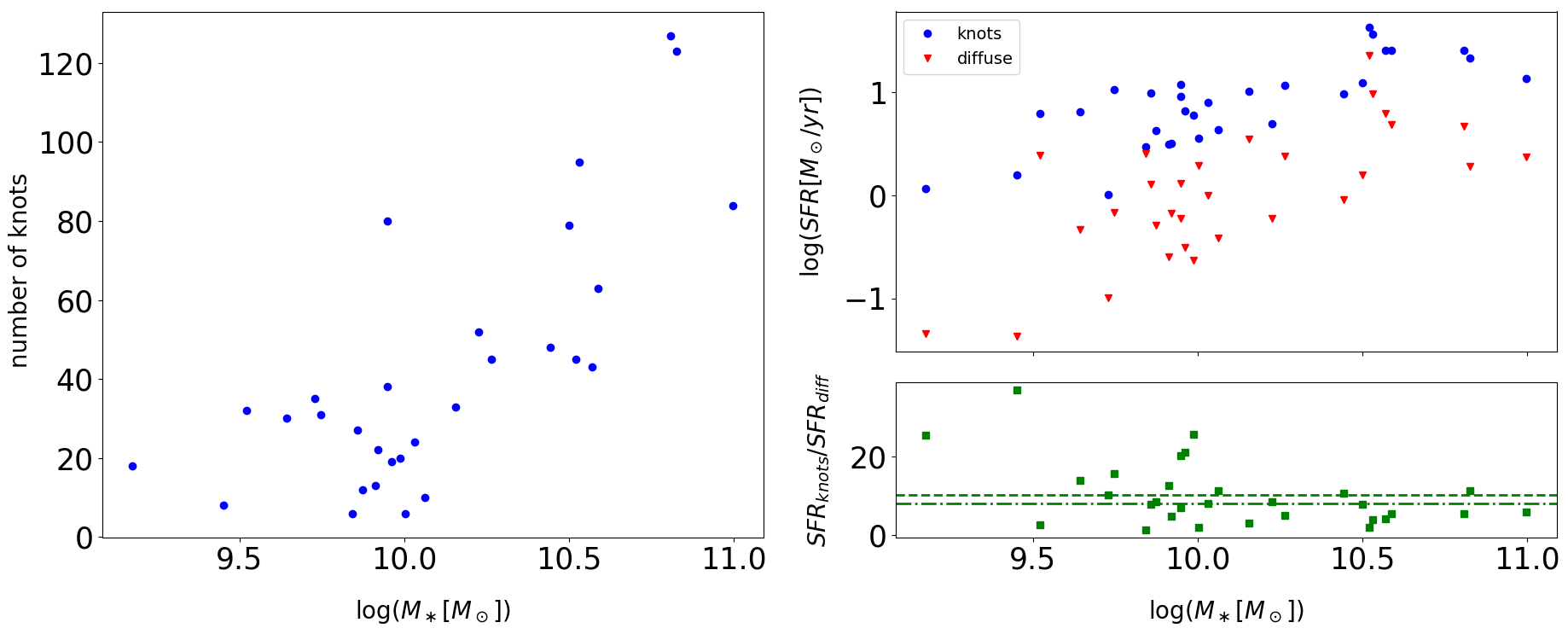}
\caption{Number of knots (left) and SFR in the knots  and in the diffuse component (right) as a function of the stellar mass, for each galaxy separately The bottom right inset shows the ratio of the SFR between the two component, along with the median (dash-dotted) and the mean value (dashed). \label{fig:nknots_mass} }
\end{figure*}

\begin{figure}
\centering
\includegraphics[scale=0.4]{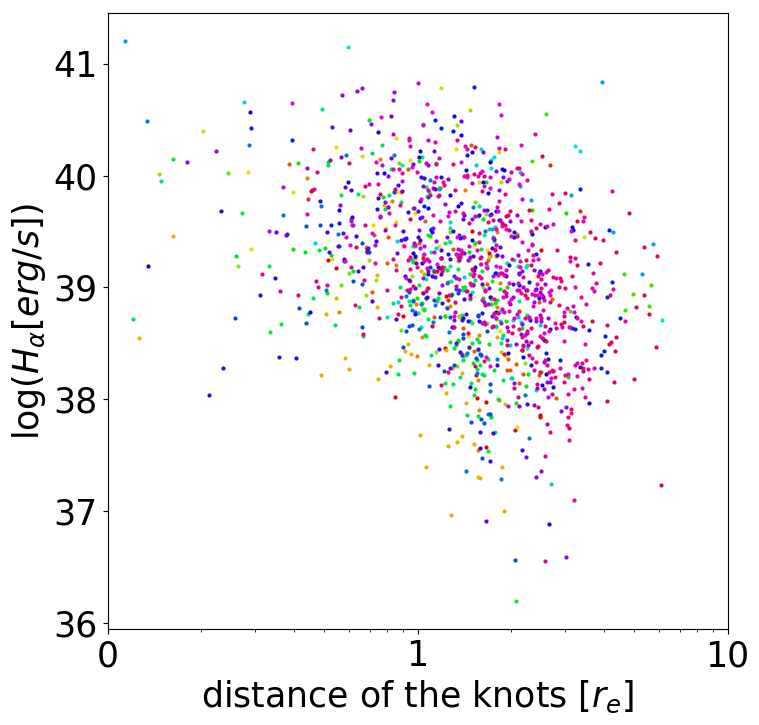}
\caption{\Ha luminosity of each knot as a function of galactocentric distance in unit of r$_e$, for all galaxies in the sample. Different colours refer to knots belonging to different galaxies. \label{fig:knots_dist} }
\end{figure}

There are several  parameters that might drive the  \Ssfr-\Sm relations. One of these is the presence and spatial distribution of concentrated star forming regions, spread throughout the galaxy disk. Indeed, Figure \ref{fig:Hamaps} shows that the all galaxies present bright \Ha knots with  \Ha surface brightness typically between $\rm 10^{-16.5}-10^{-15} \,  erg \, s^{-1} \,  cm^{-2} \,  arcsec^{-2}$. As discussed also in \citetalias{Poggianti2017a, Poggianti2019}, these are star-forming clumps embedded in regions of more diffuse emission. \citetalias{Poggianti2017a} describes in detail how these clumps are identified. Briefly, a shell script including IRAF and FORTRAN routines has been developed. The script searches the local minima of the laplace + median filtered \Ha MUSE image. The boundaries of these clumps (i.e. their radius, having assumed circular symmetry) is estimated considering outgoing shells until the average counts reach a threshold value that defines the underlying diffuse emission.

Both the number of knots and the total SFR in these knots are somehow correlated with the total mass of the galaxy, even though a quite spread exist, as shown in Fig. \ref{fig:nknots_mass} \citepalias[see also][]{Poggianti2019}. The more massive the galaxy the higher the number of knots.
The right panel of the Figure shows that the SFR in the knots is systematically higher than the SFR in the diffuse component, on average by a factor of $\sim 7$.

Figure \ref{fig:knots_dist} shows the  \Ha luminosity of the knots as a function of the galactocentric distance in units of $r_e$, for each galaxy separately. Even though an overall anticorrelation exists, with more distant knots being less \Ha luminous (and therefore less star forming), the relation spans over three orders of magnitude at a fixed galactocentric distance, showing how knots at the same galactocentric distance can have systematically different star forming properties. 

\begin{figure*}
\centering
\includegraphics[scale=0.35]{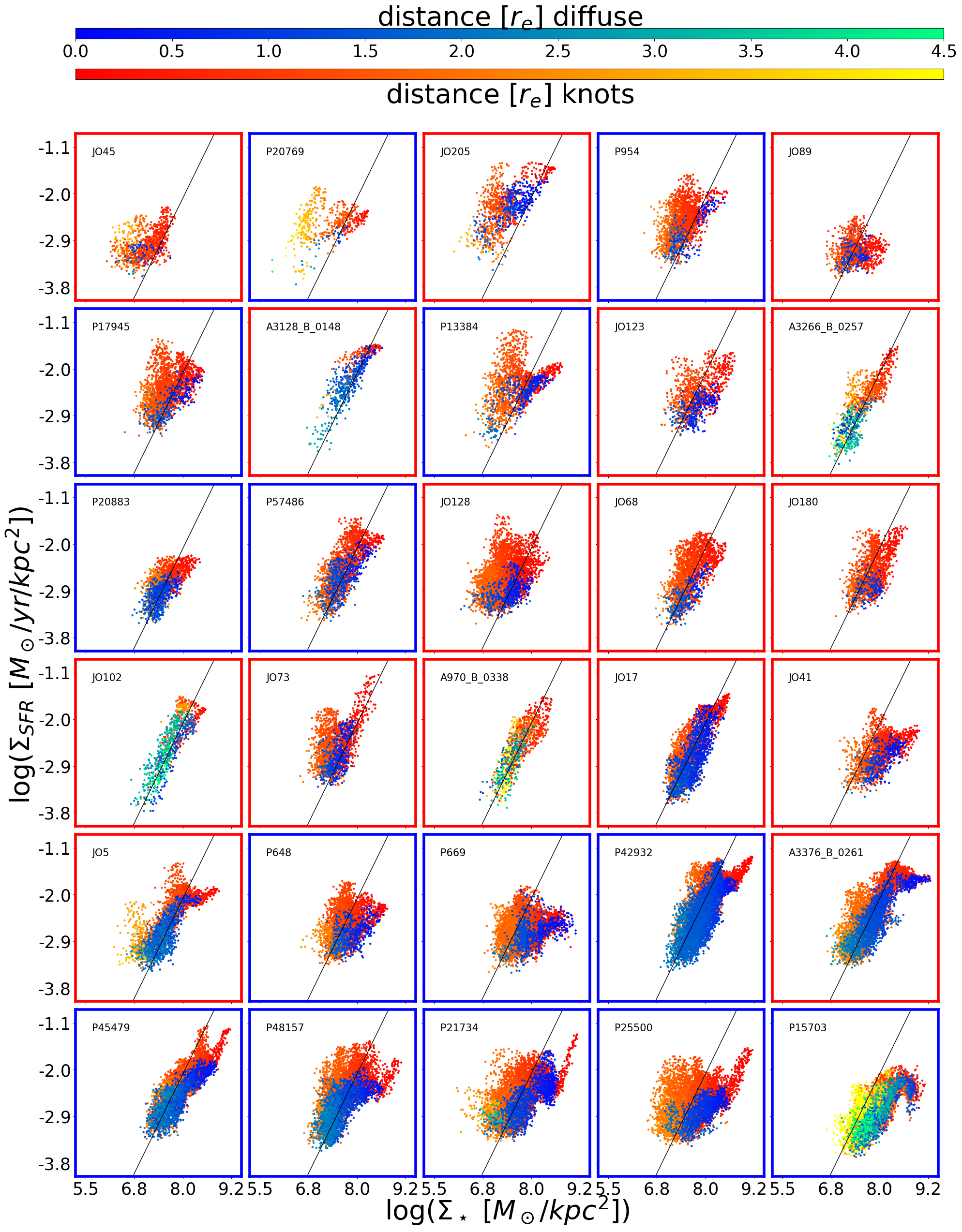}
\caption{\Ssfr - \Sm relation for all galaxies in the sample, sorted by increasing stellar mass and colour coded by the distance, in unit of $r_e$. The diffuse component of the galaxies (removing the \Ha knots) is plotted using a blue-green colour-scale, the knots are plotted using a red-yellow colour-scale. Galaxies surrounded by a red square belong to clusters, galaxies surrounded by a blue square belong to the field. The black line represents the fit to the whole sample, from Fig.\ref{fig:sfr_mass_all}. \label{fig:sfr_mass_nknots_diffuse_knots} }
\end{figure*}

\begin{figure}
\centering
\includegraphics[scale=0.38]{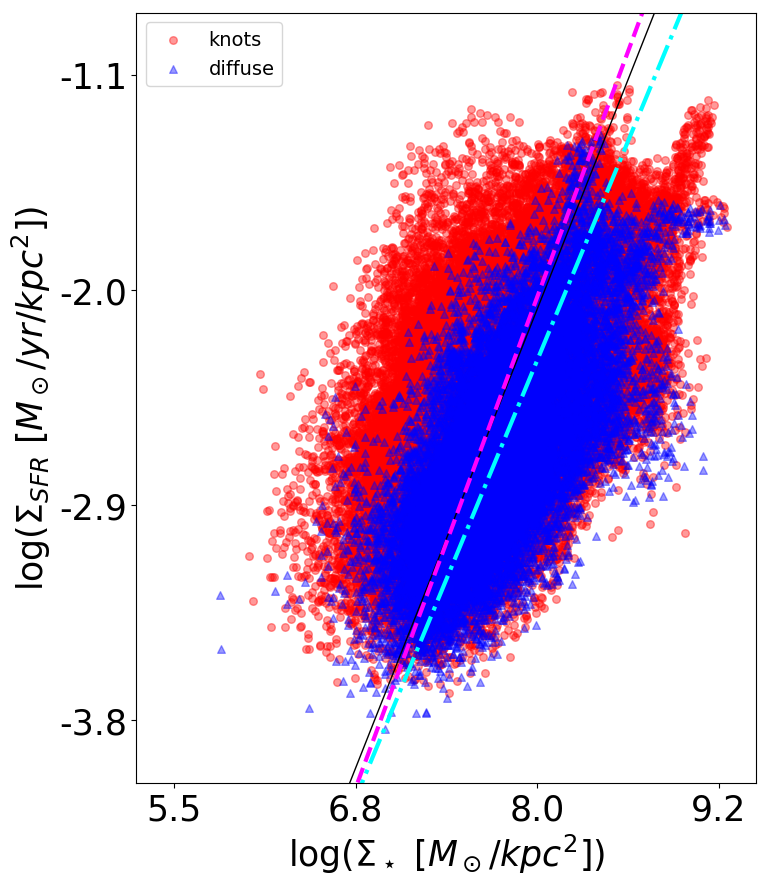}
\caption{\Ssfr - \Sm relation for spaxels belonging to knots (red) and to the diffuse component for all galaxies in the sample together. The black line represents the fit to the whole sample, from Fig.\ref{fig:sfr_mass_all}, the dashed magenta line is the fit to the knots, the dash-dotted cyan line is the fit to the diffuse component.  \label{fig:sfr_mass_diffuse_knots_all} }
\end{figure}

Figure \ref{fig:sfr_mass_nknots_diffuse_knots} shows the same relation shown in Fig.\ref{fig:sfr_mass_dist_mass}, but using different colour scales for spaxels in the knots and in the diffuse component: spaxels belonging to the bright knots are plotted using a red-yellow colour scale, spaxels belonging to the diffuse component are plotted using a blue-green colour scale. In total, 69861 of the spaxels belong to the knots, the remaining 22159 spaxels to the diffuse component.  It is immediately clear that the \Ssfr-\Sm relations for spaxels within the knots are very broad: at any given \Sm a large range of \Ssfr values can exist. In addition, there is not always a tight correlation between galactocentric distance, \Sm and \Ssfr in the knots: highly  star forming knots are not only present in the center of the galaxies and they not always take place in the highest surface mass density regions.
The multi-sequences visible in the plots are due to highly star forming knots distributed at different galactocentric distances, even at low surface mass densities. Note that if we reduced the size of the knots by half, to overcome the issue of their external regions being influenced by the diffuse component due to the size of the PSF, we obtain the same result.

\begin{figure}
\centering
\includegraphics[scale=0.35]{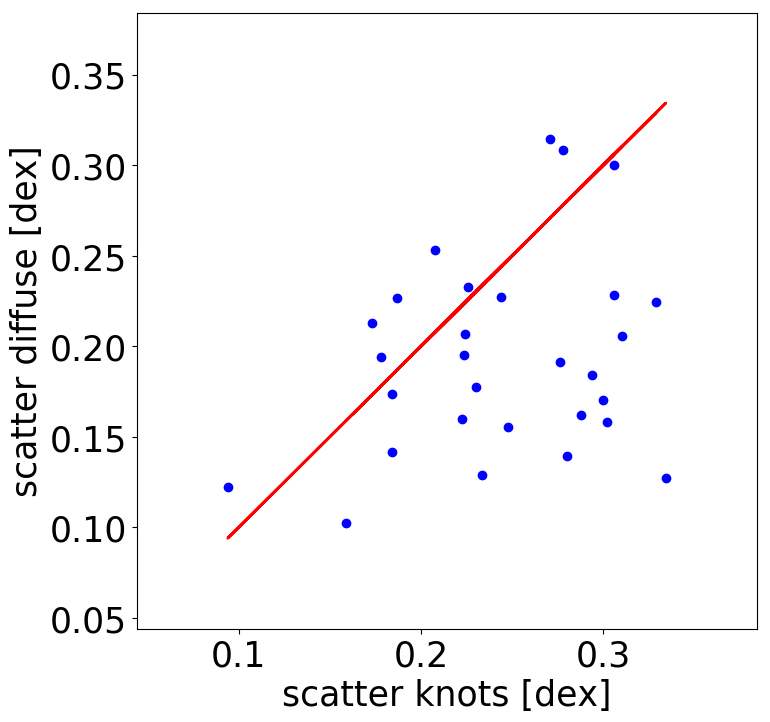}
\caption{For each galaxy, comparison between the median scatter of the \Ssfr-\Sm relation obtained using the spaxels in the knots and that obtained using those in the diffuse component. The red line represents the 1:1 relation. \label{fig:scatter_dk} }
\end{figure}

In contrast,  relations  for the diffuse component are much tighter, though still a broadening of the relations is visible for some cases.  This is even clearer in Fig.\ref{fig:sfr_mass_diffuse_knots_all}, where all galaxies are plotted together. It is interesting to note that, even though the scatter is much larger for spaxels in the knots than for the diffuse component, the slope and intercept of the relations are remarkably similar. We note that this result is not driven by the larger number of spaxels in the knots: we randomly extracted from the knots sample 10 subsamples with the same number of elements as the diffuse sample, finding the same result. 

Figure \ref{fig:scatter_dk} compares the median scatter of the \Ssfr-\Sm relation obtained using the spaxels in the knots and that obtained using those in the diffuse component, highlighting how the latter is systematically lower for more than 70\% of the sample. 

Our result suggests that  bright star forming regions follow different \Ssfr-\Sm relations from the diffuse emission.

\subsection{The \Ssfr-\Stg relation}\label{sec:gas}

\begin{figure*}
\centering
\includegraphics[scale=0.35]{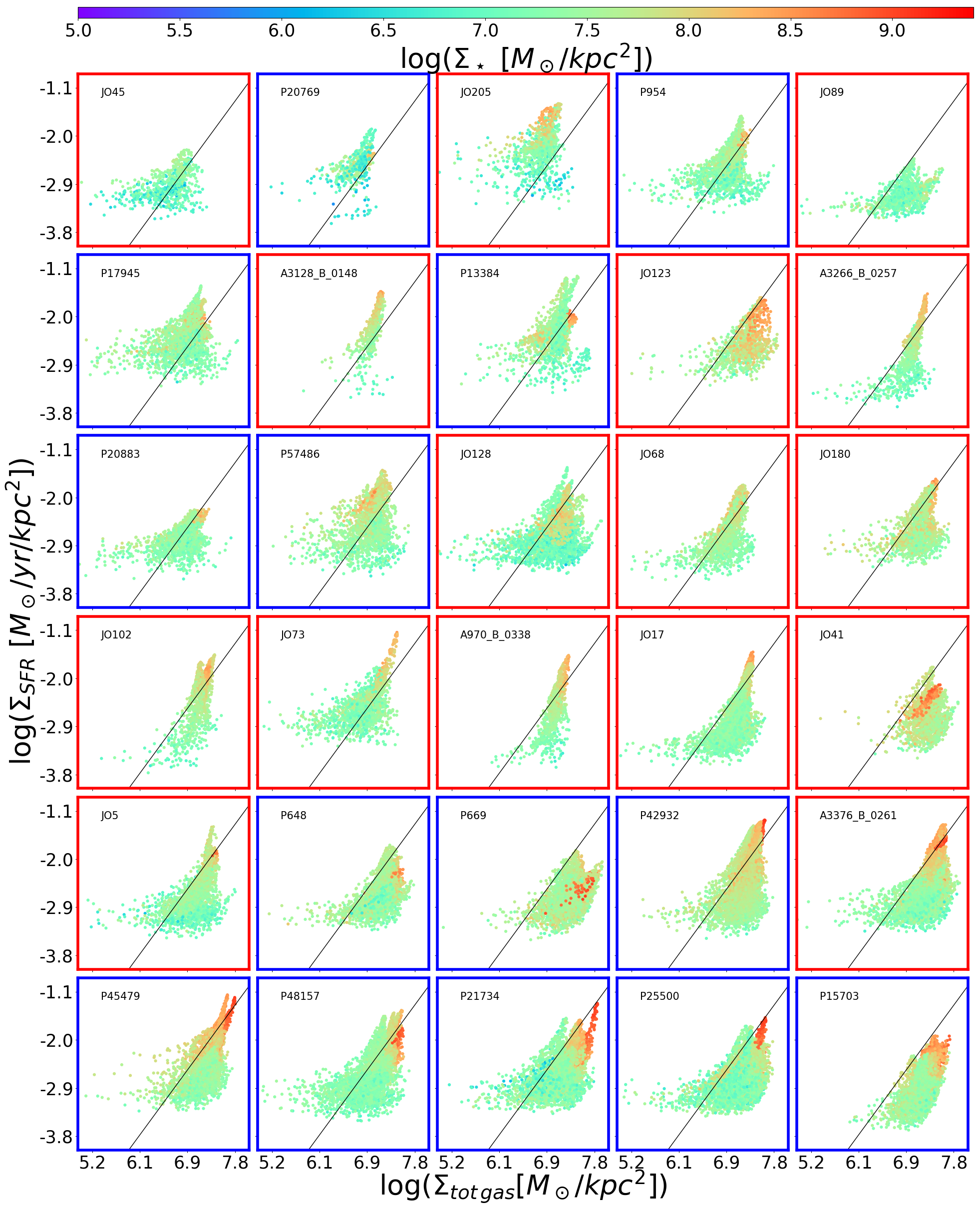}
\caption{Spatially resolved SFR-total gas mass relation for the knots and the diffuse component together for  all galaxies in the sample, sorted by total stellar mass and colour coded by the \Sm. Black line represent the  Kennicutt-Schmidt law \citet{Kennicutt1998b}. Galaxies surrounded by a red square belong to clusters, galaxies surrounded by a blue square belong to the field. \label{fig:sfr_totgasmass} }
\end{figure*}

The molecular gas surface density has been found to trace the stellar mass surface density on kiloparsec scales \citep{Lin2017}. This might lead to the correlation between \Ssfr and \Sm, at a fixed star formation efficiency (SFE). Here we therefore aim at understanding whether the \Ssfr  correlates better to the total gas surface density (\Stg) than  to the \Sm, for both the diffuse component and the knots.  

 \cite{Bigiel2008} have investigated the \Ssfr-\Stg relationship, at sub-kpc resolution, in a sample of 18 nearby galaxies, using high-resolution H{\sc i} data from The H{\sc i} Nearby Galaxy Survey \citep[THINGS,][]{Walter2008}, CO data from HERACLES \citep{Leroy2008} and the BIMA Survey of Nearby Galaxies \citep[SONG,][]{Helfer2003}. Their analysis was confined to the optical radius $r_{25}$, i.e., where the B-band magnitude drops below 25 mag arcsec$^{-2}$. Almost all star formation occurs within this radius, although they found that the H{\sc i} often extends much beyond  $r_{25}$.
They found that overall  the relationship has a scatter of $\sim$0.3 dex, but the distribution of points in the \Stg-\Ssfr parameter space varies from galaxy to galaxy, indicating that there is no universal behavior. In some cases, a single power law relates total gas and SFR over many orders of magnitudes in gas surface density. In other cases, they found a wide range of SFRs at almost the same gas column. The two quantities seem to become essentially uncorrelated where $\rm \Sigma H${\sc i} $ > \rm \Sigma H_2$. This transition typically occurs at $r>0.5 r_{25}$. Their analysis therefore suggests that a universal total gas Schmidt law \citep{Kennicutt1998b} does not exist at all galactocentric distances. According to their interpretation, galaxies with well-defined total gas Schmidt laws (and low H{\sc i}  surface densities) may have lost diffuse H{\sc i} not associated with star formation in interactions.

Figure \ref{fig:sfr_totgasmass} shows the \Ssfr-\Stg relationship for the GASP sample. We remind the reader that we obtained the \Stg from the $A_v$,using a calibration based on CO observations as we  have at our disposal direct estimates of neither H{\sc i} nor H$_2$. Note however that the $\rm A_v$ traces the dust heated by star formation and not the the cold dust emission and \cite{Corbelli2012} have  shown that the CO mass, proxy for the H$_2$ mass, is tightly correlated to the latter and not to the former. 

Similarly to what found by \cite{Bigiel2008}, we observe a large spread between the two quantities and different behaviours among the different galaxies. Typically, relations are thinner at high \Ssfr values and they widen at low \Ssfr values. Multi sequences are observed at high \Ssfr values. These ones corresponds to different knots that are characterised by different \Sm (as shown by the colour of the points) and \Stg. Some galaxies lie on the  Kennicutt-Schmidt law (even though with a large scatter - e.g. JO68), other lie well below it (e.g. P15703). The  scatter ranges from 0.2 dex (JO89) to 0.5 dex (P42932). Considering separately the knots and the diffuse component, no significant differences are observed (plot not shown). 

As \cite{Bigiel2008} found that the region at which the two quantities are essentially uncorrelated is at  $r\sim 0.5 r_{25}$, we investigate if also in our sample selecting only the central regions thins the relation. 
In our sample  $r_e \sim 0.5 r_{25}$ (G. Fasano 2019, private communication). Considering only spaxels with  $r< r_e$ (plot not shown), we obtain a scatter of $\sim0$.3 dex, compatible with what found by  \cite{Bigiel2008}. Our results therefore on one side validate the approach of using the dust extinction to compute total gas surface density, at least in the galaxy cores, and on the other side support the hypothesis that the  Kennicutt-Schmidt law  does not hold in the galaxy outskirts.  

Finally, our results also  suggest the the total gas surface density is not a primary driver of the \Ssfr-\Sm relation, as the \Ssfr-\Stg relation is not tighter than the former, when all spaxels are considered.  

\subsection{Spatially resolved vs Global SFR-Mass relation}\label{sec:global}
In the previous sections we have observed that the spatially resolved SFR-\ma relation seems not to depend on the total stellar mass. Plots were sorted by increasing stellar mass but no clear trends were visible.

\begin{figure*}
\centering
\includegraphics[scale=0.35]{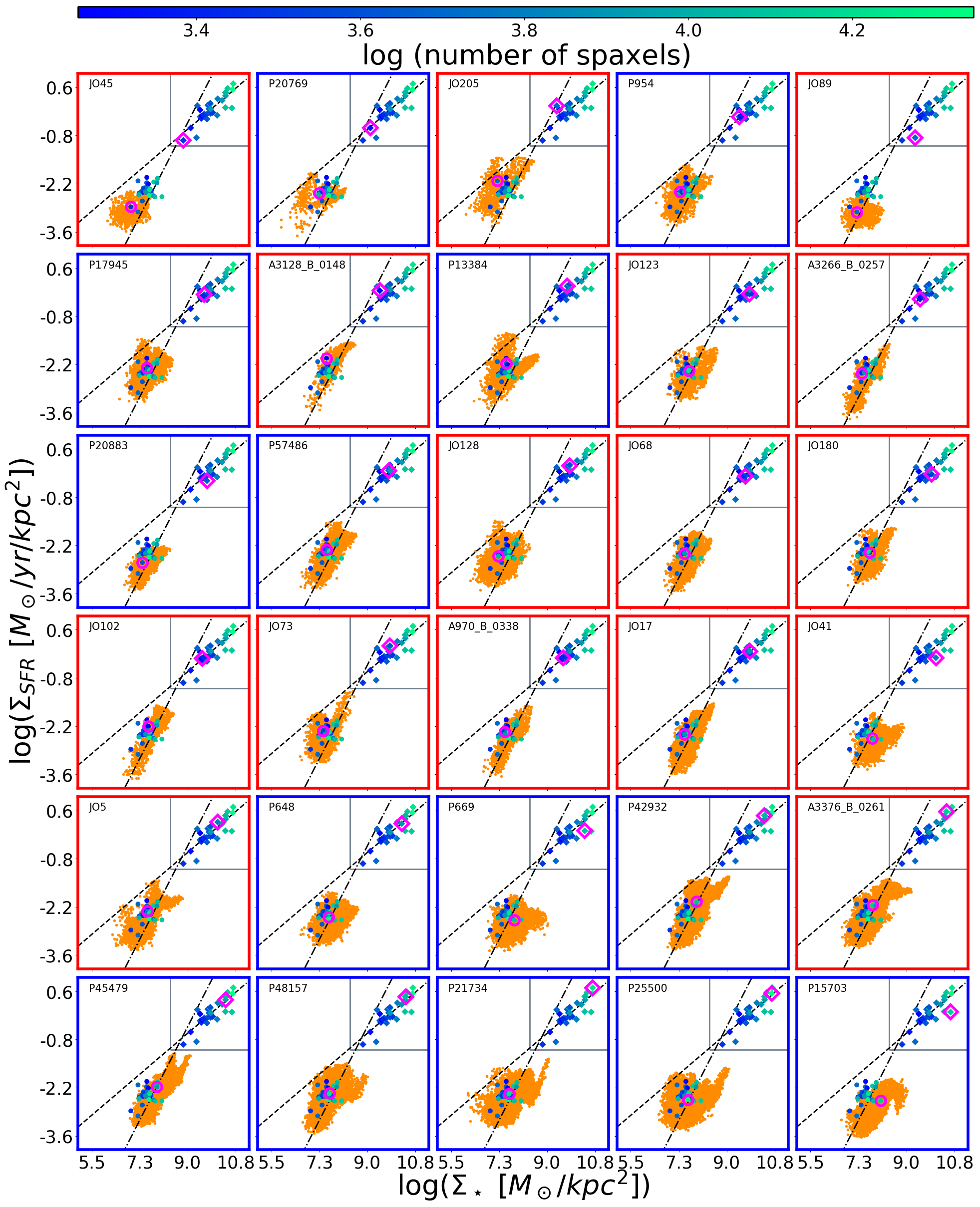}
\caption{Comparison between the local and the global SFR-\ma relations. In each panel, the \Ssfr-\Sm relation of the galaxy listed in the upper corner is shown with orange dots. The mean \Ssfr and \Sm values of the same galaxy is shown with an empty magenta circle. Filled circles show the mean \Ssfr and \Sm values for all the galaxies. They are colour coded by the number of spaxels in each galaxy and are all the same in all the panels. The axis aligned sub-panels show the global relation, obtained by summing the values of all the spaxels in each galaxy. The x-axis shows the $\log(M_\ast [M_\odot])$, the y-axis shows the $\log (SFR [M_\odot \, yr^{-1}])$ Filled diamonds are colour coded by the number of spaxels in each galaxy, the empty magenta diamond indicates the position of the galaxy shown in that panel.  In all the panels,  the dash-dotted line  shows the fit from the upper left panel of Fig.\ref{fig:sfr_mass_all}, the dashed line shows the fit to the global SFR-Mass relation. Galaxies surrounded by a red square belong to clusters, galaxies surrounded by a blue square belong to the field. \label{fig:sfr_mass_gl_res} }
\end{figure*}

It is now interesting to understand how the local relation is related to the global one. Fig. \ref{fig:sfr_mass_gl_res} resembles Fig. \ref{fig:sfr_mass_dist_mass}, where each panel shows the \Ssfr-\Sm relation for each galaxy separately. Single spaxels are represented with orange colours. Each panel also shows the mean \Ssfr and \Sm value for 
each galaxy of the sample, represented by coloured circles. The
circles are colour coded by  the number of spaxels for which the stellar mass has been computed, used as a proxy of the galaxy size, in the corresponding galaxy. The circle surrounded by an empty magenta one indicates the values of the galaxy shown in that panel. Each panel also shows the global SFR-\ma relation of each galaxy, obtained by summing the values of all the spaxels of a given galaxy. The relation is shown with diamond symbols. As for the mean  \Ssfr-\Sm relation, symbols are colour coded by the number of spaxels in that galaxy and the empty magenta diamond shows the galaxy discussed in that panel.

 Figure \ref{fig:sfr_mass_gl_res} shows that overall, the mean \Ssfr and \Sm  values occupy a well defined and quite restricted region in the plane. 90\% of the points have \Ssfr values spanning the range $(6.2-1.2)\times 10^{-2}$\msyk  and \Sm values spanning the range $(1.6-8.2)\times 10^7$ \msk.
 They therefore span a range of $\sim 0.7 dex$ both in surface mass density and in surface SFR density. Hence even though each galaxy is characterised by a different and overall quite spread \Ssfr-\Sm relation, average values are similar. No trends are seen for the mean \Ssfr and \Sm values with increasing number of spaxels within the galaxies, indicating that these quantities are independent on the galaxy size. In contrast, a clear trend with galaxy size is seen when global values are considered: bigger galaxies (indicated by lighter colours) are found in the upper left region of the global SFR-\ma relation, while smaller galaxies (indicated by darker colours) are preferentially  located in the bottom left region.  In addition, the ranges spanned by global values are almost twice as  broad as those spanned by local values: SFR values range from 0.12 to 5 $M_\odot \, yr^{-1}$, stellar mass values from 8$\times 10^8$ to 5$\times 10^{10}$ $M_\odot$.  The relation is therefore stretched and presents a flatter fit than the fit of the local relation. 
 This finding indicates that the global SFR-Mass relation arises simply from the combination of two factors: a) the mean spatially resolved \Ssfr vs \Sm value does not vary much for most galaxies, and b) more massive galaxies are on average bigger (the mass-size relation), i.e. have more spaxels that contribute both to the total mass and total SFR. This stretches the relation between the two quantities into a linear and quite tight global SFR-mass relation. Thus, the global SFR-Mass relation is a consequence of the mass-size relation. 

\section{Discussion and summary}\label{sec:disc}

In this paper we have reported the analysis of the spatially resolved SFR-\ma relation, based on 92020 star forming spaxels of  30 local undisturbed late-type galaxies  in different environments drawn from the GAs Stripping Phenomena in galaxies with MUSE \citepalias[GASP,][]{Poggianti2017a}. 

The GASP survey exploits the exquisite performances of  the  integral-field spectrograph MUSE mounted at the VLT and one of its added values  is that the observations cover the entire optical extension of the galaxy, so that we can characterise the star forming properties up to several effective radii. Specifically, we detected \Ha out of more than 2.7$r_e$ for 75\% of the sample analysed in this work.

Given the MUSE capabilities and observational conditions, we could recover the spatially resolved SFR-\ma relation on the kpc scale, so we are in the regime where SFR and \ma estimates are reliable. Indeed, while reaching sufficiently small spatial 
scales is critical to shed light on the physical processes that regulate galaxy formation, it is important to stress that measurements for the SFMS may  be appropriate only on scales larger than individual giant molecular clouds (GMCs), typically of the order of 100pc. On  smaller scales, young clusters might drift from their parent GMC, thus disconnecting measurements of \Ssfr from their related $\Sigma_{gas}$, therefore yielding incorrect \ma and SFR. 

Star formation  activity can also vary from cloud to cloud, and since most SFR transformations are generalized to be applied across all environments, sampling of statistically robust and extended star formation activity is therefore required.
The smallest region size that will contain both a complete sampling of star forming environments, IMFs used, and is not affected by the possible drift  of young stellar clusters  from their parent GMC  is 500 pc \citep{Kruijssen2014}. 
This 
spatial scale must  be used as the minimum scale over which to apply \ma and SFR transformation, to produce physically meaningful measurements.

In this work, we have selected star forming regions according to the [OI] BPT diagram. This diagram is typically  capable to highlight a contribution from other physical processes rather than star formation. Mechanisms such as thermal conduction from the surrounding hot ICM and turbulence are particularly relevant for the origin of the diffuse emission \citepalias[see, e.g.,][]{Poggianti2019}. Ours is therefore 
 a conservative choice. Nonetheless,  Appendix \ref{app:NII} shows that all the results hold also when using the [NII] diagnostic diagram.

The first result of our analysis is that, considering all galaxies together, a correlation between the \Ssfr and \Sm exists, though it is quite broad, having a scatter of 0.3 dex.  The correlation gets steeper and shifts to higher \Sm values when external spaxels are excluded  from the analysis and at a given galactocentric distance moving from less to more massive galaxies. The broadness of the relation suggests galaxy-by-galaxy variations. We note that the sample includes only late-type star forming galaxies, therefore  differences can not be driven by the morphology, as found for example by CALIFA studies \citep[e.g.][]{GonzalezDelgado2014b}. { No conclusion about the role of morphology in driving the} \Ssfr-\Sm { relation can be obtained with this sample.}

The second result is that inspecting individual galaxies, each object is characterized by a distinct \Ssfr-\Sm relation, { indicating that no universal relation exists}. Some galaxies show quite elongated sequences, some other much broader. { In half of the cases the coefficient of correlation states there is no correlation between \Ssfr and \Sm, suggesting that fitting relations is meaningless.} {The steep slope of the global relation  is therefore probably just a consequence of the different intercepts of the galaxies in the sample, which casts additional doubts on its physical meaning.}

For some galaxies the scatter { - if meaningful-} depends on the \Sm, for some others multi sequences co-exists. The scatter of the relation does not depend on global parameters, such as total stellar mass,  SFR,   $\rm \Delta (SFR)$, galaxy inclination,  number of knots,  SFR in the knots,  SFR in the knots divided by the total stellar mass, $r_e$. 

 
 The third result is that the broadening of the relation and the multi-sequences are mainly due to the presence of off-center bright star forming knots, that do not always follow the relation traced by the diffuse component.  
This result might reflect the stochasticity of the knots distribution: GMCs, that are regions where the knots form, do not follow a uniform or gaussian distribution, but assemble and condense "randomly" throughout the disk. SFR is not directly connected to the mass in stars in the corresponding area: indeed the mass is the integral of the star formation occurred in that region during the entire galaxy life, while star formation is the result of the ongoing gas availability. 

We note that our results differ from  those presented by \cite{Maragkoudakis2017}. They  measured the SFR and \ma of the bright off-nuclear regions and calculated the slope and dispersion of the produced \Ssfr-\ma relations, finding that higher surface brightness regions are not the primary drivers of the correlation for all spaxels. The discrepancy might arise from the different techniques adopted to select bright knots. \cite{Maragkoudakis2017}, indeed, only focused on the 10 brightest regions on the galaxy disk of each galaxy, without applying any cut in surface brightness, as we do. 

 Considering the star-forming knots and diffuse component separately, we have found that the SFR in the knots is 7$\times$ higher than in the diffuse component. The radial distribution of the knots is very scattered. As a consequence, the \Ssfr-\Sm relation for the knots is much broader than that of the diffuse component. 
The fraction of star formation happening within bound clusters \citep[Cluster Formation Efficiency, CFE][]{Bastian2012} has been investigated in a number of other single galaxies \citep{Silva-Villa2013, Adamo2015} and it has been found that it instead varies positively with both the SFR density and SFR. Galaxies with higher SFR also have on average a larger CFE \citep[see, e.g.,][]{Goddard2010, Adamo2011, Ryon2014}. 
Differences might be due to the resolution scale reached by the different works.

Another important result of our analysis is that the \Ssfr-\Stg gas relation is as broad as the \Ssfr-\Sm relation, indicating that the surface gas density is not a primary driver of the relation.

Finally,  we have  tried to reconcile the local \Ssfr-\Sm relation to the global one, which we studied in \citetalias{Vulcani2018c}. In that and in many other studies \citep[e.g.,][just to cite a few]{Brinchmann2004, Salim2007, Noeske2007a, Noeske2007b}, it has been found that on global scales SFR and \ma are tightly related, suggesting that the overall current gas content is related to the current stellar mass.

We have found that the global SFMS is a continuous relation extended from the local one \citep[see also][]{Hsieh2017, CanoDiaz2016}, but it spans a much wider area on the plane.  While on local scales the mean \Ssfr and \Sm values for all galaxies are quite similar,  with mean  \Ssfr and \Sm values varying of at most 0.7 dex across galaxies, regardless of the galaxy size, on global scales more extended galaxies are also more massive and more star-forming, suggesting that the SFR-\ma relation might be driven by the size-mass relation.

\section*{Acknowledgements}
We thank the referee, Yago Ascasibar, for his constructive report that helped us to improve the manuscript. Based on observations collected at the European Organisation for Astronomical Research in the Southern Hemisphere under ESO programme 196.B-0578. This project has received funding from the European Reseach Council (ERC) under the Horizon 2020 research and innovation programme (grant agreement N. 833824).  We acknowledge funding from the INAF PRIN-SKA 2017 program 1.05.01.88.04 (PI Hunt). We acknowledge financial contribution from the contract ASI-INAF n.2017-14-H.0.  Y.~J. acknowledges support from CONICYT PAI (Concurso Nacional de Inserci\'on en la Academia 2017) No. 79170132.






\bibliographystyle{mnras}
\bibliography{gasp} 




\appendix

\section{The \Ssfr-\Sm relation obtained using the [NII]-based diagnostic diagram}\label{app:NII}

In the main text we have employed the diagnostic diagram based on the [OI] line ([OIII]5007/$\rm H\beta$ vs [OI]6300/$\rm H\alpha$)
to distinguish between regions powered by star formation and regions powered by other mechanisms. In this Appendix we instead adopt the [NII]-based diagnostic diagram  ([OIII]5007/$\rm H\beta$ vs [NII]6583/$\rm H\alpha$) and show that results are robust again this choice.

To separate the regions powered by Star-formation, Composite (SF+LINER/AGN), AGN and LINER emission we adopt the division lines by \cite{Kauffmann2003b, Kewley2001, Sharp2010}.

Using the  [NII] diagram, 142354 spaxels are powered by star formation. The star-forming sample is therefore 1.5$\times$ larger than that discussed in the main paper. Nonetheless, also the [NII] diagram does not detect any  AGN in the sample. 99391 spaxels belong to knots, 42963 belong to the diffuse component. This selection therefore slightly increases the incidence of the diffuse component. 
The vast majority of spaxels excluded according the [OI] diagram, but included when adopting the [NII] diagram are characterised by low values of both \Ssfr and \Sm. 

Considering all galaxies together, the \Ssfr-\Sm relation has a scatter of 0.35 dex, but it is qualitatively very similar to what presented in Fig. \ref{fig:sfr_mass_all} (plot not shown).

\begin{figure*}
\centering
\includegraphics[scale=0.35]{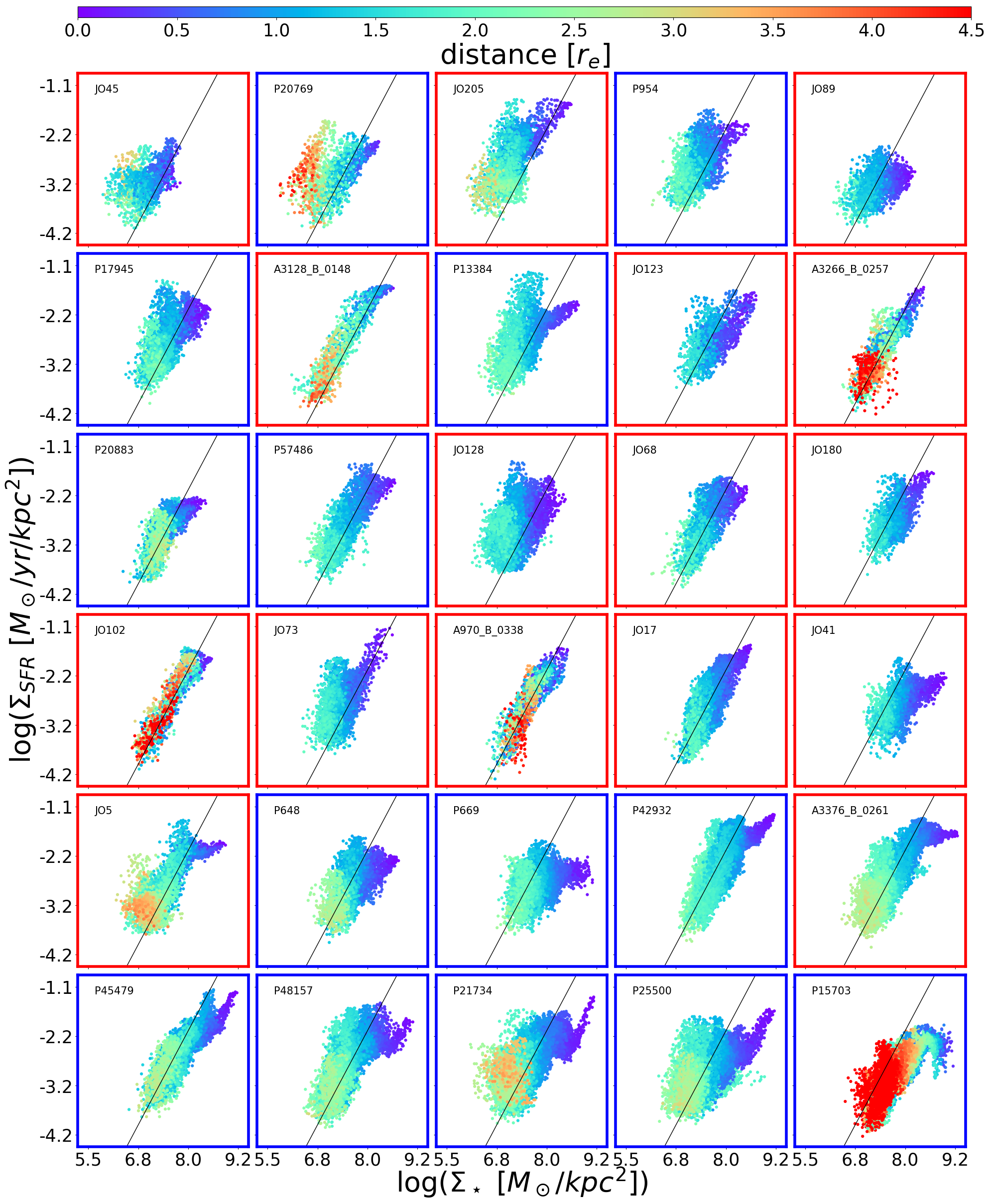}
\caption{\Ssfr - \Sm relation for all galaxies in the sample, sorted by their stellar mass and colour coded by the galactocentric distance of each spaxel, in unit of $r_e$. Regions powered by star formation according to the [NII]-based diagnsotic diagram are used.  Galaxies surrounded by a red square belong to clusters, galaxies surrounded by a blue square belong to the field. The black line represents the fit to the whole sample. \label{fig:sfr_mass_dist_mass_NII} }
\end{figure*}

Figure \ref{fig:sfr_mass_dist_mass_NII} shows the \Ssfr-\Sm relation for each galaxy of the sample separately, similarly to what shown in Fig.\ref{fig:sfr_mass_dist_mass}. 
Besides the larger number of spaxels plotted, the \Ssfr-\Sm relations are very similar to the ones presented in  the main text. They extend more towards lower values of both \Ssfr and \Sm values, but these stay on the same relation traced by the other spaxels. 

We have also checked that  any of the results presented in the paper depends on the choice of the diagnostic diagram adopted, we can therefore conclude that the choice does not affect our findings.

\section{The galaxy-by galaxy variation of the relation}\label{app:scatter}

\begin{figure*}
\centering
\includegraphics[scale=0.3]{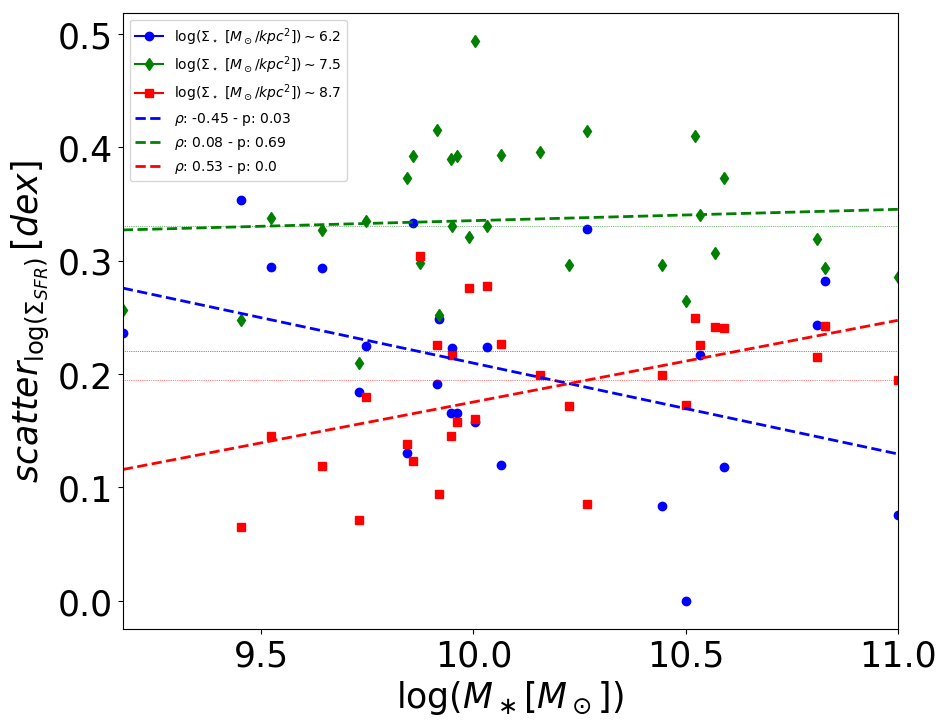}
\includegraphics[scale=0.3]{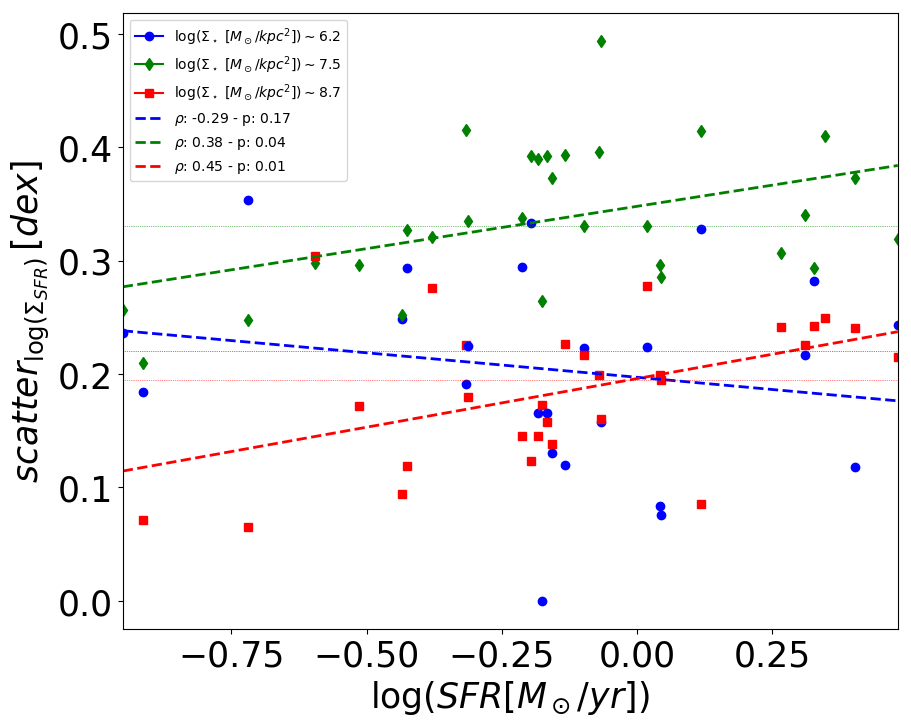}
\includegraphics[scale=0.3]{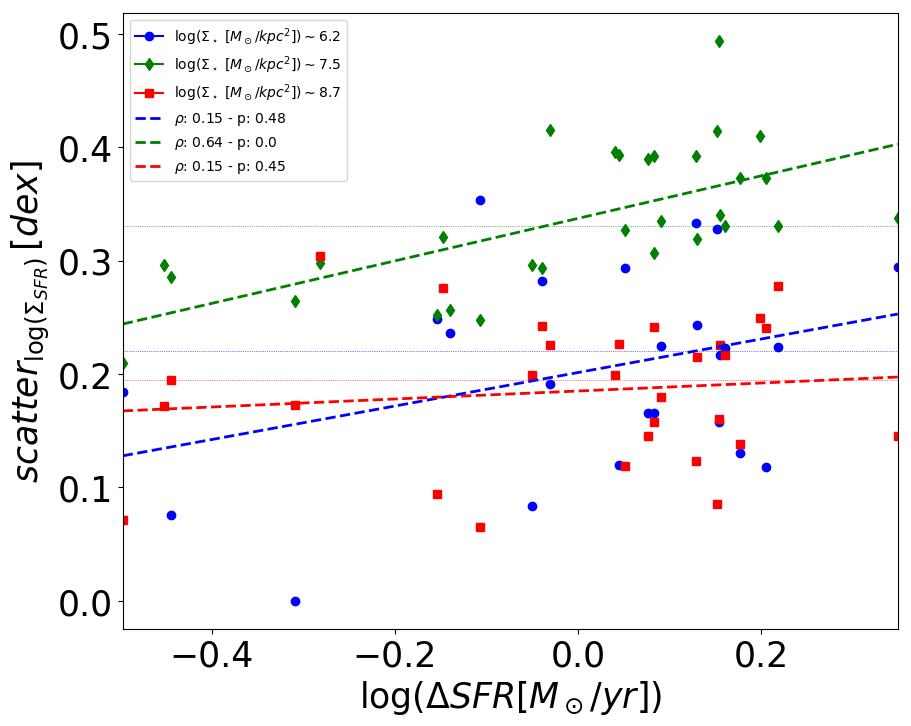}
\includegraphics[scale=0.3]{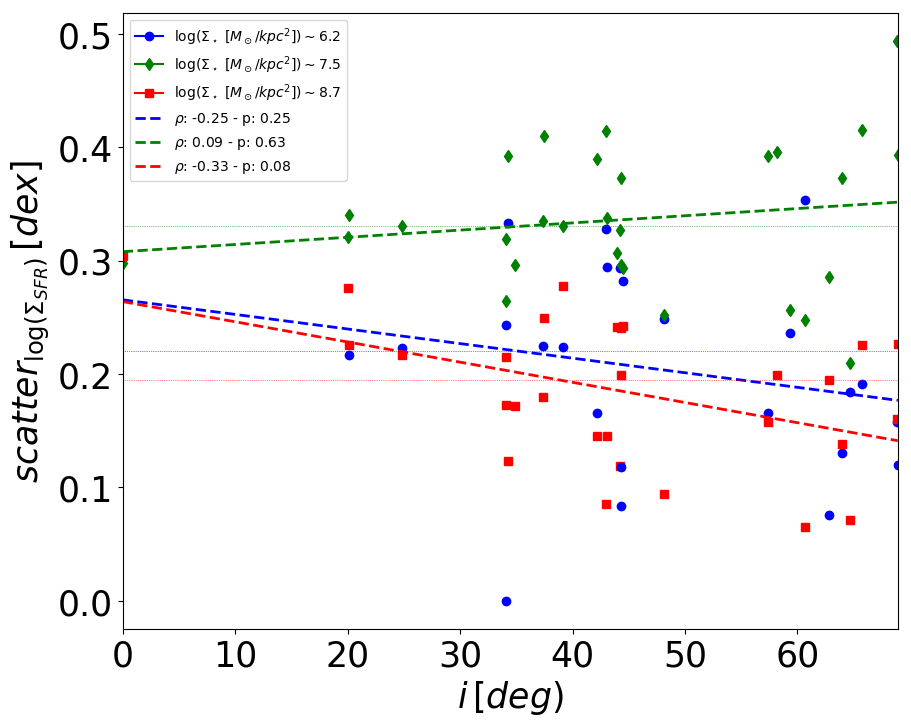}
\includegraphics[scale=0.3]{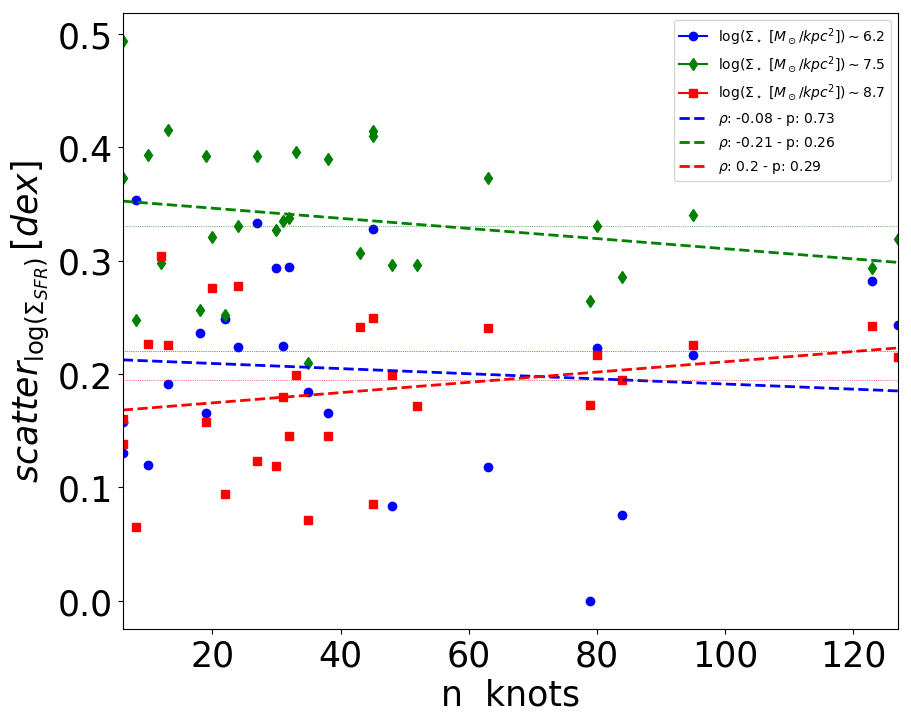}
\includegraphics[scale=0.3]{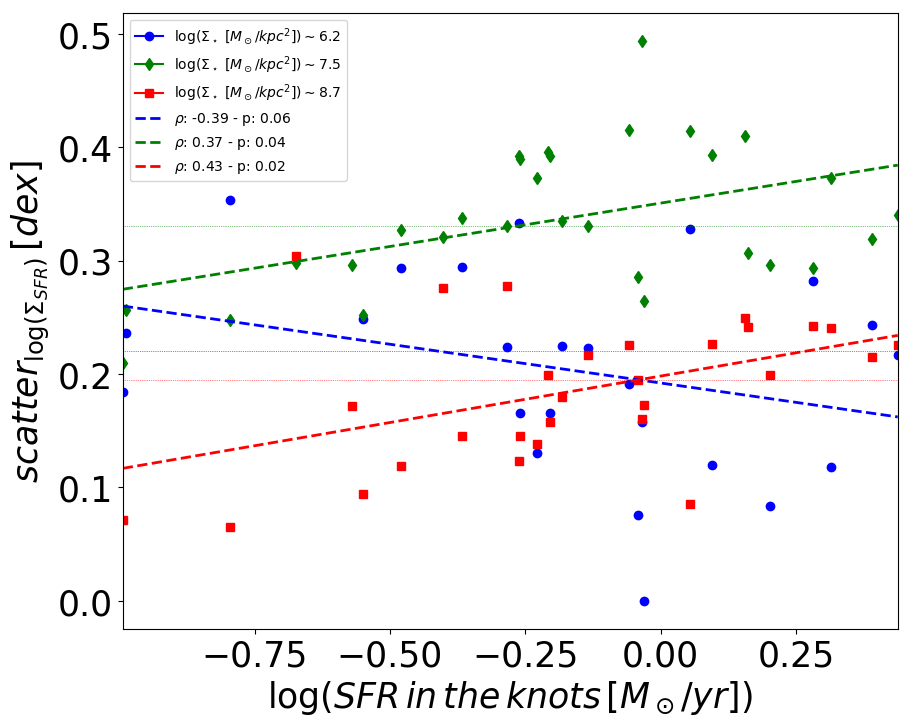}
\includegraphics[scale=0.3]{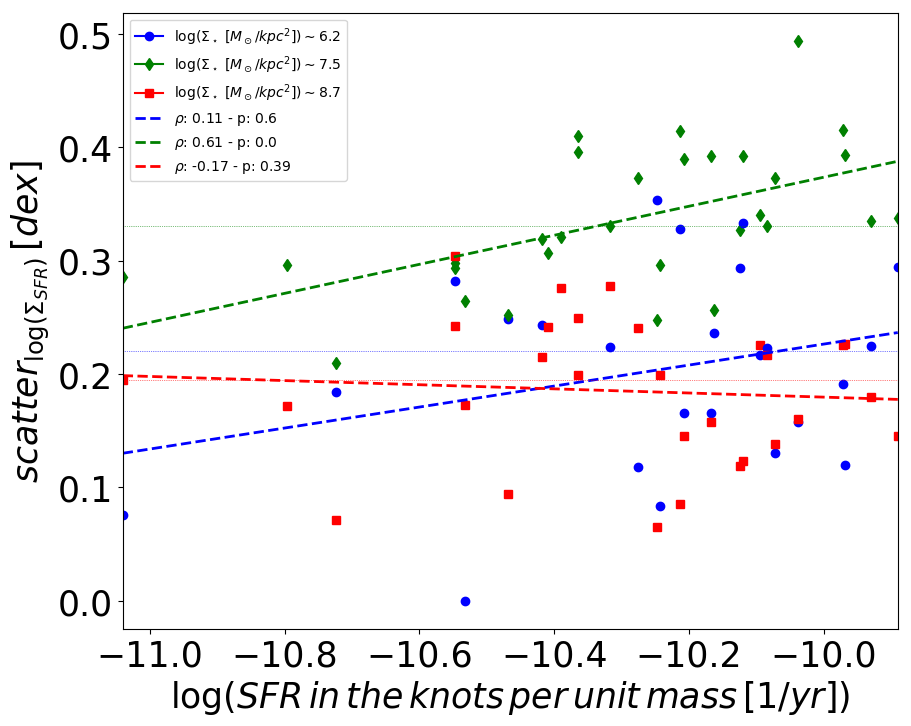}
\includegraphics[scale=0.3]{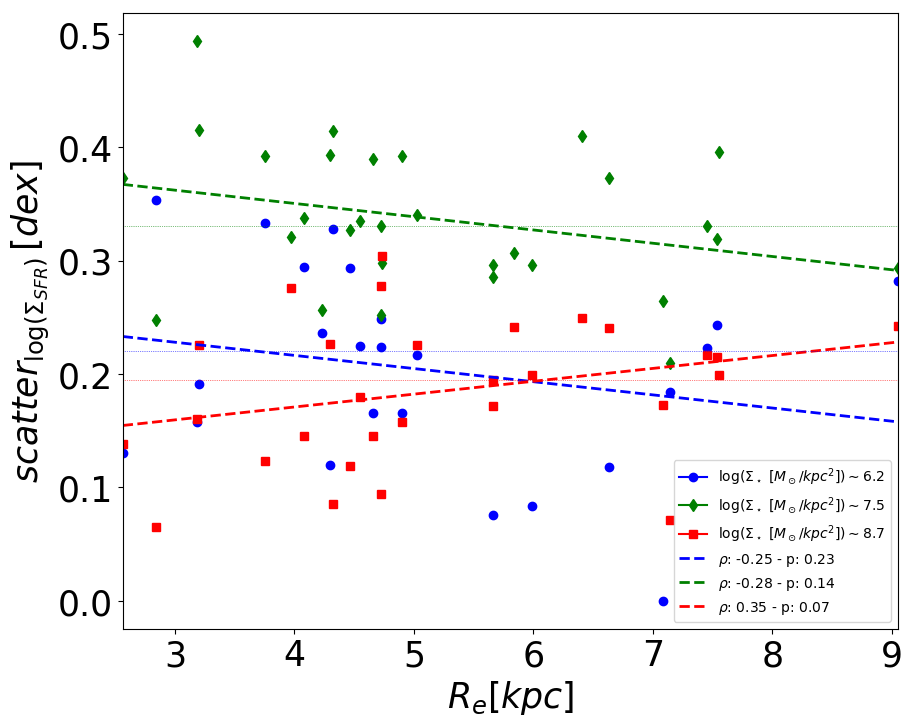}
\caption{Scatter of the median \Ssfr in three different \Sm bins, for all galaxies in the sample, as a function of different parameters. From top to bottom, left to right: the total stellar mass, the SFR,  the $\rm \Delta (SFR)$ (see text for details), the galaxy inclination, the number of knots, the SFR in the knots, the SFR in the knots divided by the total stellar mass, the effective radius. Light horizontal lines show the median values in the three bins, dashed thick lines show the linear fits to the sets of points. In the legend, the correlation coefficients and p-values of the Spearman correlation test are given. \label{fig:scatter}} 
\end{figure*}

The magnitude of the scatter of the \Ssfr-\Sm relation might depend on a number of global parameters. Figure \ref{fig:scatter}
shows this quantity measured in three surface mass density bins as a function of the galaxy total stellar mass, the SFR,  the $\rm \Delta (SFR)$ - i.e. the differences between the galaxy SFRs and their expected value according to the fit of the SFR-\ma relation, given their mass \citepalias{Vulcani2018c}, the galaxy inclination, the number of knots, the SFR in the knots, the SFR in the knots divided by the total stellar mass, the effective radius, respectively. We run the nonparametric Spearman test to  measure the monotonicity of the relationships. 
The correlation coefficient varies between -1 and +1, with 0 implying no correlation. No strong and statistically meaningful correlations are found for any of the parameters analysed, as either the correlation coefficient $\rho$ is close to zero or the p-value is $>0.05$. The only positive correlation supported by the statistical test is that with $\Delta (SFR)$ in the intermediate surface mass density bin.

\bsp	
\label{lastpage}
\end{document}